\def\eop{\mathcal{E}}
\def\bop{\mathcal{B}}
\def\e{\hat{\bf e}}
\def\sraise{\;\raise1.0pt\hbox{$'$}\hskip-6pt\partial}
\def\slower{\;\overline{\raise1.0pt\hbox{$'$}\hskip-6pt\partial}}
\def\threej#1#2#3#4#5#6{\left( \begin{array}{ccc} #1 & #2 & #3 \\ #4 & #5 & #6 \end{array} \right) }
\def\eqdef{\stackrel{\rm def}{=}}
\def\Tr{\mbox{Tr}}
\def\be{\begin{equation}}
\def\ee{\end{equation}}
\def\bea{\begin{eqnarray}}
\def\eea{\end{eqnarray}}
\def\nn{\nonumber}
\def\bx{{\bf x}}
\def\bigoh{{\mathcal O}}
\def\qtwo{\qquad\qquad}
\def\qthree{\qquad\qquad\qquad}
\def\qfour{\qquad\qquad\qquad\qquad}
\def\qfive{\qquad\qquad\qquad\qquad\qquad}
\def\qsix{\qquad\qquad\qquad\qquad\qquad\qquad}

\def\simle{\lesssim}
\def\simge{\gtrsim}

\documentclass[twocolumn,amsmath,amssymb,prd,showpacs]{revtex4}
\usepackage{bm}
\usepackage{epsf}

\newcommand{\aut}[2]{{#2.\ #1,}}
\newcommand{\paut}[2]{{#2.\ #1} and}
\newcommand{\laut}[2]{{#2.\ #1,}}
\newcommand{\etal}{{\it et al. }}

\newcommand{\miscj}[4]{#1, {\bf #2}, #3 (#4).}
\newcommand{\ApJ}[3]{Astrophys.\ J., {\bf #1}, #2 (#3).}
\newcommand{\ApJS}[3]{Astrophys.\ J.\ Supp., {\bf #1}, #2 (#3).}
\newcommand{\ApJL}[3]{Astrophys.\ J.\ Lett., {\bf #1}, #2 (#3).}
\newcommand{\MNRAS}[3]{Mon.\ Not.\ R.\ Astron.\ Soc., {\bf #1}, #2 (#3).}
\newcommand{\PRD}[3]{Phys.\ Rev.\ D, {\bf #1}, #2 (#3).}
\newcommand{\PRL}[3]{Phys.\ Rev.\ Lett., {\bf #1}, #2 (#3).}
\newcommand{\pp}[1]{preprint, astro-ph/#1.}
\newcommand{\book}[2]{#1, #2.}
\newcommand{\astroph}[1]{}

\begin{document}

\title{Pseudo-$C_\ell$ estimators which do not mix E and B modes}
\author{Kendrick M. Smith}
\affiliation{Kavli Institute for Cosmological Physics, University of Chicago, 60637}

\begin{abstract}
\baselineskip 11pt
Pseudo-$C_\ell$ quadratic estimators for CMB temperature and polarization power
spectra have been used in the analysis pipelines of many CMB experiments, such 
as WMAP and Boomerang.  In the polarization case, these estimators mix E and B 
modes, in the sense that the estimated B-mode power is nonzero for a noiseless 
CMB realization which contains only E-modes.  Recently, Challinor \& Chon 
showed that for moderately sized surveys ($f_{sky} \sim 0.01$), this mixing 
limits the gravity wave B-mode signal which can be detected using pseudo-$C_\ell$ 
estimators to $T/S \sim 0.05$.  We modify the pseudo-$C_\ell$ construction, 
defining ``pure'' pseudo-$C_\ell$ estimators, which do not mix E and B modes 
in this sense.  We study these estimators in detail for a survey geometry 
similar to that which has been proposed for the QUIET experiment, for a variety 
of noise levels, and both homogeneous and inhomogeneous noise.  For noise levels
$\simle 20$ $\mu$K-arcmin, our modification significantly improves the B-mode 
power spectrum errors obtained using pseudo-$C_\ell$ estimators.  In the
homogeneous case, we compute optimal power spectrum errors using a Fisher
matrix approach, and show that our pure pseudo-$C_\ell$ estimators are roughly
80\% of optimal, across a wide range of noise levels.  There is no limit,
imposed by the estimators alone, to the value of $T/S$ which can be detected.
\end{abstract}


\maketitle

\section{Introduction}
In the past few years, polarization of the cosmic microwave background anisotropy at the 10\% level, 
a long-standing prediction of Big Bang cosmology \citep{BEpol},
has been detected by several experiments \citep{DASI1,CBI1,CAPMAP,Boom03}.
The strongest detection to date excludes zero at the $10\sigma$ level \citep{CBI2}.
Going beyond first detection, a primary goal for CMB experiments in the next decade will be making precision measurements of polarization power spectra.
In addition to providing a strong consistency check for the standard $\Lambda$CDM cosmological model, 
and improving existing uncertainties on cosmological parameters,
this will break parameter degeneracies which cannot be resolved using CMB temperature alone \citep{ZSS}.
Examples include 
constraining the reionization history of the universe \citep{Zreion,HH},
reconstructing the dark matter distribution at $z\sim 1$ via the lensing B-mode \citep{OHdelens,HSdelens},
breaking degeneracies among isocurvature modes \citep{BMT},
and measuring the primordial fluctuations \citep{HOpp,YW}.

One of the most tantalizing prospects for CMB polarization experiments is the possibility of detecting the B-mode signal
from primordial gravity waves \citep{SZgw,KKSgw}.
In the standard $\Lambda$CDM model, B-modes offer a unique observational window on these waves, since there is no source
of gravity waves (tensor perturbations) after inflation, and scalar perturbations generate only E-modes in linear perturbation theory.
The only cosmological contaminants are therefore higher-order effects such as gravitational lensing of the E-mode spectrum \citep{ZSlens}.

On a practical level, many methods have been proposed for estimating power spectra from CMB polarization data.
A fully optimal, likelihood-based anaylsis scales as $\bigoh(N_{pix}^3)$ \citep{BJK}, and will probably not be feasible for next generation experiments.
An alternative method, which has become the ``industry standard'' for CMB temperature experiments with $N_{pix} \simge 10^5$, is using pseudo-$C_\ell$
quadratic estimators \citep{WHGCl}.
In the polarization context, pseudo-$C_\ell$ estimators mix E into B, in the following sense:
for a noiseless CMB realization containing only E-modes, the estimated B-mode bandpowers will be nonzero.
In an ensemble of such realizations, the estimated B-mode bandpowers will be zero in the mean (this is because
pseudo-$C_\ell$ estimators are unbiased by construction), but because they are nonzero in each realization,
E-mode signal power does contribute to the variance of the B-mode estimators.
This extra variance can be thought of as a source of noise which is due to the estimators alone.
Recently, \citet{ChalChon} showed that this ``estimator noise'' can dominate the sample variance from lensing B-modes, so
that it becomes the dominant contaminant if the instrumental noise is sufficiently small.
For surveys with $f_{sky}\sim 0.01$, they show that it limits the value of $T/S$ which can be detected to $\sim 0.05$.

The purpose of this paper is to construct pure pseudo-$C_\ell$ estimators; these are
modified versions of polarization pseudo-$C_\ell$ estimators, which do not mix E into B
in the sense defined above.  Using these estimators, the estimated B-mode power will be zero for any noiseless CMB realization
which contains only E-modes.
Strictly speaking, this is only true in the continuum limit; in a finite pixelization, the B-mode estimators will acquire small
nonzero values from pixelization artifacts.  We show that these pixelization effects can be made arbitrarily small by increasing
the resolution.

The basic idea of our construction was inspired by the pure B-mode formalism of \citep{LCT,BunnPure}.
Because pure B-modes exist in any finite patch of sky, a BB power spectrum estimator will receive no contribution from EE signal power if it
is built entirely out of these modes.
Our pure pseudo-$C_\ell$ estimators have the property that the observed polarization map always appears contracted with a pure B-mode
which is constructed from heuristically chosen weight functions.
The main technical difficulty is ensuring that the fast algorithm for calculating the transfer matrix \citep[][Appendix E]{HGpolCl}
still goes through after this modification.

Throughout this paper, we use a fiducial $\Lambda$CDM cosmology which is consistent with WMAP \citep{WMAPcp}, with parameters
\{ $\Omega_b h^2$, $\Omega_m h^2$  $\Omega_\Lambda$,   $\tau$,      $|\Delta R|^2$,    $n_s$,  $w$, $m_\nu$ \} = 
 \{     0.024,          0.14,         0.73,       0.17,  $2.57\times 10^{-9}$,   1,  -1,   0 \}.
For pixelized all-sky maps, we use the Healpix coordinate system \citep{HPIXobligatory,HPIXwebsite} exclusively.
We use a phenomonological definition of $T/S$, defining it to be the ratio of {\em temperature} multipoles
$C^{tensor}_{\ell=10} / C^{scalar}_{\ell=10}$.

This paper is organized as follows.  
In \S\ref{se}, we briefly review pseudo-$C_\ell$ estimators, before constructing pure pseudo-$C_\ell$ estimators in \S\ref{sc}.
In \S\ref{sx1}, we consider a spherical cap shaped mock survey with uniform white noise, using it to illustrate general features of our
estimators, and show that our modification significantly improves the performance of the estimators for noise levels $\simle 20$ $\mu$K-arcmin.
As a first step toward more realistic instrumental noise, in \S\ref{sx2} we consider a survey with inhomogeneous, but not spatially correlated, noise
and show that the same conclusions apply.
In \S\ref{sts}, we study a range of noise levels and show that our estimators are roughly 80\% of optimal, defined by the degradation in
the value of $(T/S)$ which can be detected, for all noise levels considered.
We conclude in \S\ref{sco}.

\section{Notation and Conventions}
\label{sn}
We represent CMB polarization as a symmetric traceless tensor $\Pi_{ab}$ as in \citet{KKS}, but we have
changed some conventions to agree with those of CMBFAST and Healpix.
The all-sky metric $g_{ab}$ and antisymmetric tensor $\epsilon_{ab}$
are given by
\be
g_{ab} = \left(\begin{array}{cc}
1 & 0 \\
0 & \sin^2\theta
\end{array} \right)
\qquad
\epsilon_{ab} = \left(\begin{array}{cc}
0 & \sin\theta \\
-\sin\theta & 0
\end{array} \right).
\ee
We define basis polarization fields
\bea
q_{ab} &=& \frac{1}{2} \left(\begin{array}{cc}
1 & 0 \\
0 & -\sin^2\theta
\end{array} \right)  \nn  \\
u_{ab} &=& \frac{1}{2} \left(\begin{array}{cc}
0 & \sin\theta \\
\sin\theta & 0
\end{array} \right).
\eea
Given two points ${\bf x}$, ${\bf x}'$, we define $X_a$ to be the vector at ${\bf x}$ 
which points away from ${\bf x}'$ along the great circle arc which connects the two, 
vector $Y_a = -\epsilon_{ab} X^b$, and symmetric traceless tensors
$Q_{ab} = (X_aX_b - Y_aY_b)/2$, $U_{ab} = (X_aY_b + Y_aX_b)/2$.
The quantities $X'$, $Y'$, $Q'$, $U'$ are defined in the same way with ${\bf x}$, ${\bf x}'$ interchanged.
Note that we always use lower case to distinguish the basis $q_{ab}$,
$u_{ab}$, which is globally defined (except at poles), from the ``two-point'' basis
$Q_{ab}$, $U_{ab}$, which is defined relative to a pair of
points ${\bf x}$, ${\bf x}'$.

$E$ and $B$ modes are defined as follows.  First, we define operators
\bea
\eop_{ab} &=&  -\nabla_a \nabla_b + \frac{1}{2}g_{ab}\nabla^2  \\
\bop_{ab} &=& \frac{1}{2}\epsilon_{ac}\nabla^c\nabla_b + \frac{1}{2}\epsilon_{bc}\nabla^c\nabla_a  \nn
\eea
which take scalar fields to symmetric traceless tensors.  $E$ and $B$ harmonics are defined by
\begin{eqnarray}
Y^E_{(\ell m)ab} &=& \frac{1}{\sqrt{(\ell-1)\ell(\ell+1)(\ell+2)}} \eop_{ab} Y_{\ell m}  \label{en1}  \\
Y^B_{(\ell m)ab} &=& \frac{1}{\sqrt{(\ell-1)\ell(\ell+1)(\ell+2)}} \bop_{ab} Y_{\ell m}.  \nn
\end{eqnarray}
We will also need some analagous definitions for the spin-1 case.
The ``two-point'' basis $X_a$, $Y_a$ has already been defined; a
global basis $x_a$, $y_a$ is defined by:
\be
x_a = \left( \begin{array}{cc} 1 & 0 \end{array} \right)
\qquad\qquad
y_a = \left( \begin{array}{cc} 0 & \sin\theta \end{array} \right).
\ee
Spin-1 harmonics, which we label G and C for ``gradient'' and
``curl'', are defined by
\begin{eqnarray}
Y^G_{(\ell m)a} &=& \frac{1}{\sqrt{\ell(\ell+1)}} \nabla_a Y_{\ell m}         \\
Y^C_{(\ell m)a} &=& -\frac{1}{\sqrt{\ell(\ell+1)}} \epsilon_{ab} \nabla^b Y_{\ell m}
\end{eqnarray}

\section{Pseudo-$C_\ell$ estimators}
\label{se}
Pseudo-$C_\ell$ quadratic estimators for polarization power spectra have been previously constructed
by \citet{HGpolCl}, who used the pseudo-$C_\ell$ formalism \citep{WHGCl}.
In this section, we briefly recall the pseudo-$C_\ell$ construction, largely for the sake of establishing notation.

One first chooses a pixel-dependent weight function $W({\bf x})$ which is zero outside the survey region.
The choice is made heuristically, but the performance of the estimators is improved (on angular scales with
signal-to-noise ratio $\simle 1$) by choosing $W({\bf x})$ to be smaller where the noise is larger, in order to downweight noisy regions.
Frequently, the weight function is also apodized near the survey boundary, in order to reduce harmonic ringing.
One then defines pseudo $E$ and $B$ multipoles by (the tildes signify ``pseudo'')
\bea
\widetilde E_{\ell m} &=& \sum_{\bf x} 2 \Pi^{ab}({\bf x}) W({\bf x}) Y^{E*}_{(lm)ab}({\bf x}) \label{ee5} \\
\widetilde B_{\ell m} &=& \sum_{\bf x} 2 \Pi^{ab}({\bf x}) W({\bf x}) Y^{B*}_{(lm)ab}({\bf x}). \nn
\eea
and pseudo power spectra by
\bea
\widetilde C_\ell^{EE} &=& \frac{1}{2\ell+1} \sum_{m=-\ell}^\ell \widetilde E_{\ell m}^* \widetilde E_{\ell m}  \\
\widetilde C_\ell^{BB} &=& \frac{1}{2\ell+1} \sum_{m=-\ell}^\ell \widetilde B_{\ell m}^* \widetilde B_{\ell m}.  \nn
\eea
It can be shown that the expectation values of the pseudo power spectra are given by
\be
\label{ee1}
\left( \begin{array}{c}
\langle \widetilde C_\ell^{EE} \rangle  \\
\langle \widetilde C_\ell^{BB} \rangle
\end{array} \right)
=
\left(  \begin{array}{cc}
K^+_{\ell\ell'} & K^-_{\ell\ell'}  \\
K^-_{\ell\ell'} & K^+_{\ell\ell'}
\end{array} \right)
\left( \begin{array}{c}
C_{\ell'}^{EE} \\
C_{\ell'}^{BB}
\end{array} \right)
+
\left( \begin{array}{c}
\widetilde N^{EE}_\ell  \\
\widetilde N^{BB}_\ell
\end{array} \right)
\ee
where $K^\pm_{\ell\ell'}$ are $\ell_{max}$-by-$\ell_{max}$ transfer matrices
and $\widetilde N^{EE}_{\ell}$, $\widetilde N^{BB}_{\ell}$ are $\ell_{max}$-by-1 vectors
which represent additive noise bias.
There is an efficient algorithm, which will be discussed in Appendix \ref{atr}, for
exactly computing the transfer matrices from the weight function.
The noise bias can be computed exactly in cases where the noise is
uncorrelated between pixels (Appendix \ref{anb}), or by Monte Carlo in the general case.

Unbiased power spectrum estimators $\widehat C_\ell^{EE}$, $\widehat C_\ell^{BB}$ can
be obtained from the pseudo power spectra $\widetilde C_\ell^{EE}$, $\widetilde C_\ell^{BB}$ by simply subtracting the noise
bias and applying the inverse of the $(2\ell_{lmax})$-by-$(2\ell_{max})$
``grand unified transfer matrix'':
\be
\label{ee2}
\left( \begin{array}{c}
\widehat C_\ell^{EE}  \\
\widehat C_\ell^{BB}
\end{array} \right)
\eqdef
\left(  \begin{array}{cc}
K^+_{\ell\ell'} & K^-_{\ell\ell'}         \\
K^-_{\ell\ell'} & K^+_{\ell\ell'}         
\end{array} \right)^{-1}
\left( \begin{array}{c}
\widetilde C_{\ell'}^{EE} - \widetilde N^{EE}_{\ell'}  \\
\widetilde C_{\ell'}^{BB} - \widetilde N^{BB}_{\ell'} 
\end{array} \right)
\ee

The preceding construction has assumed that the power spectrum is estimated at every multipole $\ell$.  
For reasons of sky coverage or signal-to-noise, it is often necessary to bin multipoles into bandpowers with $\Delta\ell > 1$.  
In this case, for each band $b$, one defines pseudo bandpowers
\be
\widetilde C_b^{EE} = \sum_\ell P_{b\ell} \widetilde C_\ell^{EE}
\qquad\qquad
\widetilde C_b^{BB} = \sum_\ell P_{b\ell} \widetilde C_\ell^{BB},
\ee
where the matrix $P$ defines the $\ell$ weighting within each bandpower estimator.
A commonly-used choice is \citep{MASTER}:
\be
\label{ee4}
P_{b\ell} = \left\{ \begin{array}{cc}
\frac{1}{2\pi} \frac{\ell(\ell+1)}{\ell_{low}^{(b+1)} - \ell_{low}^{(b)}}  &
  \mbox{if $\ell_{low}^{(b)} \le \ell < \ell_{low}^{(b+1)}$}              \\
0 & \mbox{otherwise.}
\end{array} \right.
\ee
One also introduces a matrix $\bar P$, which defines an interpolation
scheme by which the signal power spectra depend on bandpowers $\Delta_b$:
\be
C_\ell^{EE} = \sum_b \bar P_{\ell b} \Delta_b^{EE}
\qquad\qquad
C_\ell^{BB} = \sum_b \bar P_{\ell b} \Delta_b^{BB}  \label{ee3}
\ee
A commonly-used choice, corresponding to piecewise flat power spectra, is:
\be
\bar P_{\ell b} = \left\{ \begin{array}{cc}
\frac{2\pi}{\ell(\ell+1)}  &  \mbox{if $\ell_{low}^{(b)} \le \ell < \ell_{low}^{(b+1)}$}              \\
0 & \mbox{otherwise.}
\end{array} \right.
\ee
The binned analogs $K^\pm_{bb'}$, $\widetilde N_b$ of the transfer matrices and noise bias vectors (Eq.\ (\ref{ee1})) are defined by:
\be
\left( \begin{array}{c}
\langle \widetilde C_b^{EE} \rangle  \\
\langle \widetilde C_b^{BB} \rangle 
\end{array} \right)
=
\left(  \begin{array}{cc}
K^+_{bb'} & K^-_{bb'}  \\
K^-_{bb'} & K^+_{bb'} 
\end{array} \right)
\left( \begin{array}{c}
\Delta_{b'}^{EE} \\
\Delta_{b'}^{BB}
\end{array} \right)
+
\left( \begin{array}{c}
\widetilde N^{EE}_b  \\
\widetilde N^{BB}_b
\end{array} \right)
\ee
and are related to the unbinned versions by:
$K^\pm_{bb'} = P_{b\ell} K^\pm_{\ell\ell'} \bar P_{\ell'b'}$,
$\widetilde N_b = P_{b\ell} \widetilde N_\ell$.
The binned analogs of the unbiased estimators (Eq.\ (\ref{ee2})) are defined by:
\be
\left( \begin{array}{c}
\widehat C_b^{EE}  \\
\widehat C_b^{BB}
\end{array} \right)
=
\left(  \begin{array}{cc}
K^+_{bb'} & K^-_{bb'}         \\
K^-_{bb'} & K^+_{bb'}  
\end{array} \right)^{-1}
\left( \begin{array}{c}
\widetilde C_{b'}^{EE} - \widetilde N^{EE}_{b'}  \\
\widetilde C_{b'}^{BB} - \widetilde N^{BB}_{b'}
\end{array} \right)
\ee
Strictly speaking, these are unbiased estimators of the bandpowers $\Delta_b$,
when the CMB power spectra are of the precise form (\ref{ee3}).

\section{Constructing pure pseudo-$C_\ell$ estimators}
\label{sc}
We now construct a modified version of the B-mode power spectrum estimator
which is ``pure'', in the sense that the estimator is identically zero in
any noiseless realization of the CMB which contains only E modes.
It is convenient to explain the construction first in the continuum limit
(\S\ref{sc1}), where sums over pixels can be replaced by integrals, and then
address implementational issues associated with finite pixelized maps (\S\ref{sc2}).

\subsection{Constructing pure pseudo-$C_\ell$ estimators 1: continuum limit}
\label{sc1}

The underlying reason why the unmodified pseudo-$C_\ell$ estimator is not
``pure'', in the sense defined above, can be understood as follows.
The pseudo B multipoles are defined by
\be
\label{ec2}
\widetilde B_{\ell m} = \int d^2x\, \sqrt{g} 2\Pi^{ab}(x) W(x) Y^{B*}_{(\ell m)ab}(x)
\ee
Since multiplication in position space mixes E and B modes,
the product $W(x) Y^B_{(lm)ab}(x)$ is a mixture of E and B,
even though $Y^B_{(lm)ab}(x)$ is a B-mode.
Therefore, $\widetilde B_{\ell m}$ receives a nonzero contribution from E-mode
power in $\Pi^{ab}$.

The basic idea of our construction is now easy to explain.
Writing out the definition of $Y^B_{(lm)}$ from Eq.\ (\ref{en1}), the definition of
$\widetilde B_{\ell m}$ above can be rewritten
\bea
\widetilde B_{\ell m} &=& \frac{1}{\sqrt{(\ell-1)\ell(\ell+1)(\ell+2)}} \times  \\
                      && \qquad \int d^2x\, \sqrt{g}  2\Pi^{ab}(x) W(x) \bop_{ab} (Y^*_{\ell m}(x)) \nn
\eea
Suppose that this definition is modified by simply moving $W(x)$ inside the operator $\bop_{ab}$:
\bea
\label{ec1}
\widetilde B^{pure}_{\ell m} &\eqdef& \frac{1}{\sqrt{(\ell-1)\ell(\ell+1)(\ell+2)}} \times  \\
                     && \qquad \int d^2x\, \sqrt{g} 2\Pi^{ab}(x) \bop_{ab} (W(x) Y^*_{\ell m}(x)) \nn
\eea
In addition, {\em suppose that the weight function $W(x)$ and its gradient
vanish on the boundary of the region}.
Then $\widetilde B^{pure}_{\ell m}$ will be identically zero in a noiseless CMB realization which
contains only E modes.
One way to see this is from the perspective of the pure B-mode formalism \citep{LCT,BunnPure};
since $(W(x) Y_{\ell m}(x))$ satisfies both Dirichlet and Neumann boundary conditions,
$\bop_{ab} (W(x) Y_{\ell m}(x))$ is a pure B-mode.
Alternately, a short, self-contained proof is given as follows.
A noiseless CMB realization which contains only E-modes can be written
$\Pi_{ab} = \eop_{ab} \phi$, for some scalar function $\phi(x)$.  Putting this
into the definition of $\widetilde B^{pure}_{\ell m}$ and integrating by parts twice, one gets
\bea
&& \frac{\sqrt{(\ell-1)\ell(\ell+1)(\ell+2)}}{2} \widetilde B^{pure}_{\ell m}   \\
        && \qquad =  \int d^2x\, \sqrt{g} \bigg[ \eop^{ab}\phi(x) \bigg]  \bigg[ \bop_{ab} (W(x) Y^*_{\ell m}(x)) \bigg]     \nn \\
        && \qquad =  \int d^2x\, \sqrt{g} \bigg[ (-\nabla^a\nabla^b + \frac{1}{2}g^{ab}\nabla^2) \phi(x) \bigg]               \nn \\
        && \qfour                 \times  \bigg[ \epsilon_{ac}\nabla^c\nabla_b  (W(x) Y^*_{\ell m}(x)) \bigg]                \nn \\
        && \qquad =  \int d^2x\, \sqrt{g} \bigg[ \epsilon_{ac}\nabla^c\nabla_b (\nabla^a\nabla^b - \frac{1}{2}g^{ab}\nabla^2) \phi(x) \bigg]  \nn \\
        && \qsix                  \times  \bigg[ W(x) Y^*_{\ell m}(x) \bigg]                                                 \nn \\
        && \qquad\qquad - \oint\! d\sigma\,\, t_a \bigg[ (-\nabla^a\nabla^b + \frac{1}{2}g^{ab}\nabla^2) \phi(x) \bigg]       \nn \\
        && \qfive                 \times  \bigg[ \nabla_b  (W(x) Y^*_{\ell m}(x)) \bigg]                                     \nn \\
        && \qquad\qquad - \oint\! d\sigma\,\, t_a \bigg[ (\nabla^a + \frac{1}{2}\nabla^a\nabla^2)\phi(x) \bigg]  W(x) Y^*_{\ell m}(x)         \nn \\
        && \qquad = 0.  \nn
\eea
Here, $\oint\! d\sigma$ denotes integration around the survey boundary and $t_a$ denotes the unit tangent vector.
The boundary terms are zero because $W(x)$ and $\nabla_a W(x)$ are assumed to vanish on the boundary.

To avoid confusion, we will denote the ``pure'' pseudo B multipole defined in Eq.\ (\ref{ec1}) by $\widetilde B^{pure}_{\ell m}$,
and denote the ``mixed'' multipole defined by Eq.\ (\ref{ec2}) by $\widetilde B^{mixed}_{\ell m}$, for the rest of this paper.
This one change, replacing $\widetilde B^{mixed}_{\ell m}$ by $\widetilde B^{pure}_{\ell m}$, is the only modification we propose to
the pseudo-$C_\ell$ formalism of \S\ref{se}.
In particular, we leave the definitions of $\widetilde E_{\ell m}$ and $\widetilde C^{EE}_\ell$ unchanged.

By analogy with the definition of $\widetilde B^{pure}_{\ell m}$, it is possible to define a ``pure E'' multipole
$\widetilde E^{pure}_{\ell m}$, but we expect that incorporating this would {\em worsen} the performance of the EE
power spectrum estimator.  This is because ambiguous modes, which are useful for measuring E-mode power,
would be discarded.  Note that we have named our estimators ``pure pseudo-$C_\ell$'' even though we are really
using a ``mixed'' estimator for E-modes and a ``pure'' estimator for B-modes.

\subsection{Constructing pure pseudo-$C_\ell$ estimators 2: finite pixelized maps}
\label{sc2}

Before moving on to construct pseudo power spectra and transfer matrices for the modified estimators,
we explain how the definition of $\widetilde B^{pure}_{\ell m}$ (Eq.\ (\ref{ec1})) is to be interpreted when $\Pi_{ab}(x)$ is a finite
pixelized map, instead of an idealized continuous field.
The technical obstacle is making sense out of the object $\bop_{ab}(W Y_{\ell m})$ which contains covariant
derivatives, and appears on the right hand side of Eq.\ (\ref{ec1}).
We first expand $\bop_{ab}(W Y_{\ell m})$ using the product rule
\bea
\bop_{ab}(f_1 f_2) &=& (\bop_{ab}f_1)f_2 + f_1(\bop_{ab}f_2)            \\
                   &&  \qquad + 2T_{abcd}(\nabla^c f_1)(\nabla^d f_2),   \nn
\eea
where we have introduced the tensor
\be
T_{abcd} = \frac{1}{4}\epsilon_{ac} g_{bd} + \frac{1}{4}\epsilon_{ad} g_{bc} + 
           \frac{1}{4}\epsilon_{bc} g_{ad} + \frac{1}{4}\epsilon_{bd} g_{ac}.
\ee
(Note that contracting a spin-2 object with $T$ results in a $45^\circ$ rotation,
e.g. $T_{abcd}q^{cd} = -u_{ab}$, $T_{abcd}\eop^{cd} = -\bop_{ab}$.)

Performing the derivatives which act on $Y_{\ell m}$, and replacing the integral by a sum over pixels,
this results in the following form for $\widetilde B^{pure}_{\ell m}$.
(We do not include the pixel area as an overall prefactor, since the normalization of $\widetilde B^{pure}_{\ell m}$
will eventually drop out when we define unbiased power spectrum estimators in Eq.\ (\ref{ec3}).
However, straightforward conversion of the integral to a sum does implicitly assume an equal-area pixelization.)
\bea
\widetilde B^{pure}_{\ell m} &=& \sum_{\bf x}  2\Pi^{ab}({\bf x}) W({\bf x}) Y^{B*}_{(\ell m)ab}({\bf x})   \label{ec4}  \\
                             &&  \qquad  + N'_\ell \sum_{\bf x} 2 T^{abcd} \Pi_{ab}({\bf x}) W_c({\bf x}) Y^{G*}_{(\ell m)d}({\bf x}) \nn \\
                             &&  \qquad  + N_\ell \sum_{\bf x} 2 T^{abcd} \Pi_{ab}({\bf x}) W_{cd}({\bf x}) Y^*_{\ell m}({\bf x})   \nn
\end{eqnarray}
where we have defined
\begin{eqnarray}
W_a      &=& \nabla_a W    \label{ec5}  \\
W_{ab}   &=& (\nabla_a \nabla_b - (1/2)g_{ab}\nabla^2) W  \nonumber \\
N'_\ell  &=& 2/\sqrt{(\ell-1)(\ell+2)}                    \nonumber \\
N_\ell   &=& 1/\sqrt{(\ell-1)\ell(\ell+1)(\ell+2)}\, .    \nonumber
\end{eqnarray}
In the form (\ref{ec4}), the only covariant derivatives are those which act on the weight function in the definitions (\ref{ec5})
of $W_a$, $W_{ab}$.
If $W(x)$ is of known analytical form (e.g., Gaussian or cosine apodization), then $W_a$, $W_{ab}$ can
simply be computed analytically.  More generally, assuming that $W(x)$ is slowly varying compared to the
pixel scale (which will be the case for any sensible weighting), $W_a$ and $W_{ab}$ can be computed from
$W(x)$ by finite differencing.  This is nontrival to implement for an irregular spherical pixelization 
such as Healpix; we present one finite differencing scheme in detail in Appendix \ref{afd}.

In either case, once a prescription for computing $W_a$, $W_{ab}$ from $W$ has been specified,
we use Eq.\ (\ref{ec4}) as the {\em definition} of $\widetilde B^{pure}_{\ell m}$ in a finite pixelization.
Note that the first term on the right-hand side is the unmodified pseudo multipole $\widetilde B^{mixed}_{\ell m}$;
the second and third terms can be thought of as counterterms which cancel the E-B mixing and involve
spin-1 and spin-2 weight functions $W_a$, $W_{ab}$.
The relative weights of the three terms are $\ell$-dependent, in a way which downweights the
counterterms at high $\ell$.
This is in qualitative agreement with the general picture of E-B mixing arising from survey
boundaries \citep{Bunn,LCT}
in which the mixing is a less significant contaminant at high $\ell$.

The definition (\ref{ec4}) suggests an efficient algorithm for evaluating the pseudo multipoles $\widetilde B^{pure}_{\ell m}$.
The first term on the right-hand side is computed, for all ($\ell$, $m$) simultaneously, using
a fast spin-2 spherical harmonic transform of the weighted polarization field $W(x) \Pi_{ab}(x)$.
The second and third terms are analagously computed using a spin-1 transform of the vector field
\bea
2 T^{abcd} \Pi_{ab} W_c &=& (\Pi_Q W_Y - \Pi_U W_X) x^d   \label{ec8}  \\
                         && \qquad  + (\Pi_Q W_X + \Pi_U W_Y) y^d  \nn
\eea
and a spin-0 (ordinary spherical harmonic) transform of the scalar function
\be
2 T^{abcd} \Pi_{ab} W_{cd} = \Pi_Q W_U - \Pi_U W_Q.   \label{ec8a}
\ee
(In Eqs.~(\ref{ec8}) and~(\ref{ec8a}), we use subscripts $X,Y$ to denote components of a
vector field in the ``global'' $\{x_a,y_a\}$ basis, and subscripts $Q,U$ to denote
components of a spin-2 field in the $\{q_{ab},u_{ab}\}$ basis.)
The prefactors $N'_\ell$, $N_\ell$ are applied after the transforms.
This algorithm computes every $\widetilde B^{pure}_{\ell m}$ from the map $\Pi_{ab}(x)$,
with the same asymptotic complexity $\bigoh(\ell_{max}^3)$ as in the
unmodified pseudo-$C_\ell$ formalism of \S\ref{se},
but the overall constant is 2-3 times worse, since one spin-0,
one spin-1, and one spin-2 harmonic transform are needed, rather than a single spin-2
transform.
Since fast spin-1 harmonic transforms are not implemented in the Healpix library, we implemented
them as part of this paper; the details are presented in Appendix \ref{avh}.

We have now constructed ``pure'' pseudo multipoles $\widetilde B^{pure}_{\ell m}$ for finite pixelized maps;
we conclude this section by defining pure pseudo power spectra and unbiased estimators of the power spectrum.
This is done in complete analogy with \S\ref{se}.
The pseudo power spectrum $\widetilde C^{BB,pure}_\ell$ is defined by
\be
\widetilde C^{BB,pure}_\ell = \frac{1}{2\ell+1} \sum_{m=-\ell}^\ell \widetilde B^{pure*}_{\ell m} \widetilde B^{pure}_{\ell m}  \label{ec9}
\ee
The expectation values of $\widetilde C^{BB,pure}_\ell$ and the unmodified pseudo spectrum $\widetilde C^{EE}_\ell$ are given by
\bea
\left( \begin{array}{c}
\langle \widetilde C^{EE}_\ell \rangle  \\
\langle \widetilde C^{BB,pure}_\ell \rangle
\end{array} \right)
&=&
\left(  \begin{array}{cc}
K^{+}_{\ell\ell'} & K^{-}_{\ell\ell'}  \\
K^{-pure}_{\ell\ell'} & K^{+pure}_{\ell\ell'}
\end{array} \right)
\left( \begin{array}{c}
C_{\ell'}^{EE} \\
C_{\ell'}^{BB}
\end{array} \right)     \nn \\
&& \qquad +
\left( \begin{array}{c}
\widetilde N^{EE}_\ell  \\
\widetilde N^{BB,pure}_\ell 
\end{array} \right).    \label{ec7}
\eea
An essential ingredient in the method is a fast algorithm for computing the new transfer matrices $K^{\pm pure}_{\ell\ell'}$.
In Appendix \ref{atr}, we present such an algorithm and show that the computational cost is $\bigoh(\ell_{max}^3)$, which
is the same as the cost of evaluating the estimators once.  (Noise bias is discussed in Appendix \ref{anb}.)
The matrix $K^{-pure}_{\ell\ell'}$ measures the contribution to $\langle \widetilde C^{BB,pure}_\ell \rangle$
by E-mode power, and is therefore expected to be zero.
Strictly speaking, this is only true in the continuum limit; in a finite pixelization, $K^{-pure}_{\ell\ell'}$ will acquire
a small nonzero value from pixelization artifacts.
This will be studied quantitatively in \S\ref{sx1}.

Finally, we define pure pseudo-$C_\ell$ estimators $\widehat C^{EE,pure}_\ell$, $\widehat C^{BB,pure}_\ell$ by
simply subtracting noise bias and applying the inverse of the $(2\ell_{max})$-by-$(2\ell_{max})$
transfer matrix:
\bea
\left( \begin{array}{c}
\widehat C^{EE,pure}_\ell  \\
\widehat C^{BB,pure}_\ell
\end{array} \right)
&=&
\left(  \begin{array}{cc}
K^+_{\ell\ell'} & K^-_{\ell\ell'}         \\
K^{-pure}_{\ell\ell'} & K^{+pure}_{\ell\ell'} 
\end{array} \right)^{-1}  \label{ec3}     \\
&& \qquad\times
\left( \begin{array}{c}
\widetilde C^{EE}_{\ell'} - \widetilde N^{EE}_{\ell'}  \\
\widetilde C^{BB}_{\ell'} - \widetilde N^{BB,pure}_{\ell'}
\end{array} \right).   \nn
\eea
Note that, even though we have not changed the definition of the pseudo spectrum $\widetilde C^{EE}_\ell$,
the definition of the unbiased estimator $\widehat C^{EE}_\ell$ has been modified in our formalism.
This is because the matrix inversion on the right-hand side of Eq.\ (\ref{ec3}) mixes all rows of the
$(2\ell_{max})$-by-$(2\ell_{max})$ matrix.
In practice, we have found that the change in $\widehat C^{EE}_\ell$ is miniscule, but we do introduce
the notation $\widehat C^{EE,pure}_\ell$ to distinguish between the two versions.

\section{An example with homogeneous noise}
\label{sx1}
In the preceding section, pure pseudo-$C_\ell$ estimators $\widehat C^{EE,pure}_\ell$, $\widehat C^{BB,pure}_\ell$,
which do not mix E and B modes have been defined.
In this section, these estimators will be studied in detail for a specific mock survey, namely a spherical ``cap''
of radius $r = 13^\circ$, with uniform white noise, and a Gaussian beam with 25' FWHM.  Parameters similar to these
were originally proposed for the QUIET experiment \cite{QUIETwebsite}, which provided the original motivation for this
paper.  (We note that since completion of this paper, the proposed QUIET beam size has changed to 10'.)
Power spectra will be estimated in ``flat'' ($C_\ell \propto 1/\ell(\ell+1)$) bands with $\Delta l = 40$.
An exception is the lowest band, where we have restricted the $\ell$ range to $10\le\ell\le 40$; we have found that
trying to estimate power at $\ell\simle 10$, which is above the survey scale, gives spurious results.

We will assess the performance of the estimators by comparing their Monte Carlo covariance with the 
Fisher matrix,
\be
\label{ex11}
F_{bb'} = \frac{1}{2} \Tr( {\bf S}_b ({\bf S}_0 + {\bf N})^{-1} {\bf S}_{b'} ({\bf S}_0 + {\bf N})^{-1} )
\ee
Here, ${\bf N}$ is the ($2N_{pix}$)-by-($2N_{pix}$) noise covariance matrix,
${\bf S}_0$ is the signal covariance matrix in the fiducial model,
and ${\bf S}_b$ is the signal covariance matrix associated to each flat bandpower in EE or BB.
The Cramer-Rao inequality asserts that the Fisher matrix is a lower bound on the covariance of
any unbiased power spectrum estimator; conversely, minimum-variance unbiased quadratic estimators
\citep{TegQE} have a covariance matrix which is equal to the Fisher matrix, but require
dense matrix computations which are prohibitively expensive for the survey sizes considered
in this paper.

The infeasibility of dense matrix algebra also means that the Fisher matrix (\ref{ex11}) cannot
be computed in a straightforward fashion.
Our method for making the calculation affordable is to exploit azimuthal symmetry of the survey region and noise.
Taking a Fourier transform in the azimuthal coordinate $\varphi$, the matrices ${\bf S}$, ${\bf N}$ will be block diagonal
in the azimuthal wavenumber $m$ ({\bf S} is still dense in the coordinate $\theta$), which makes the matrix operations in 
Eq.\ (\ref{ex11}) affordable.
Details of the method are presented in Appendix \ref{afi}.
We note that this method also could be used to make optimal power spectrum estimation affordable,
but it applies only for surveys in which both the sky coverage and noise are azimuthally symmetric.
Such surveys therefore permit benchmark comparisons between pseudo-$C_\ell$ estimators and
optimal, likelihood-based methods,
for survey sizes which are large enough that comparison would normally be infeasible.

To use the pseudo-$C_\ell$ method, one must heuristically choose a weight function $W(x)$.
We will make a modest effort to optimize this by hand, deferring a systematic framework for a future paper.
The only possibility considered for $W(x)$ will be cosine apodization,
\be
\label{ex12}
W(\theta,\varphi) = \left\{ \begin{array}{cc}
                           1                                &   \mbox{for $\theta \le r-r_*$}   \\
\frac{1}{2} - \frac{1}{2}\cos\left( \pi \frac{r-\theta}{r_*} \right)   &   
                                          \mbox{for $r-r_* \le \theta \le r$}  \\
                           0                                &   \mbox{for $\theta \ge r$}
\end{array} \right.
\ee
The parameter $r_*$ is an apodization length which will be chosen shortly.
Note that this choice of $W(x)$ satisfies the requirement of \S\ref{sc}: both $W$ and its gradient vanish at the survey boundary.
We obtain the spin-1 and spin-2 weights $W_a$ and $W_{ab}$ (Eq.\ (\ref{ec5})) by analytic calculation, rather than finite differencing:
\begin{eqnarray}
W_a &=&
  -\frac{\pi}{2r_*} \sin\left( \pi \frac{r-\theta}{r_*} \right) x_a   \\
W_{ab} &=& \left[
 \frac{\pi^2}{2r_*^2} \cos\left( \pi \frac{r-\theta}{r_*} \right)
+\frac{\pi\cot(\theta)}{2r_*} \sin\left( \pi \frac{r-\theta}{r_*} \right)
\right] q_{ab}   \nonumber
\end{eqnarray}                       
for $r-r_* \le \theta \le r$, and 0 otherwise.

\begin{figure*}
\centerline{ \epsfxsize=3.0truein\epsffile{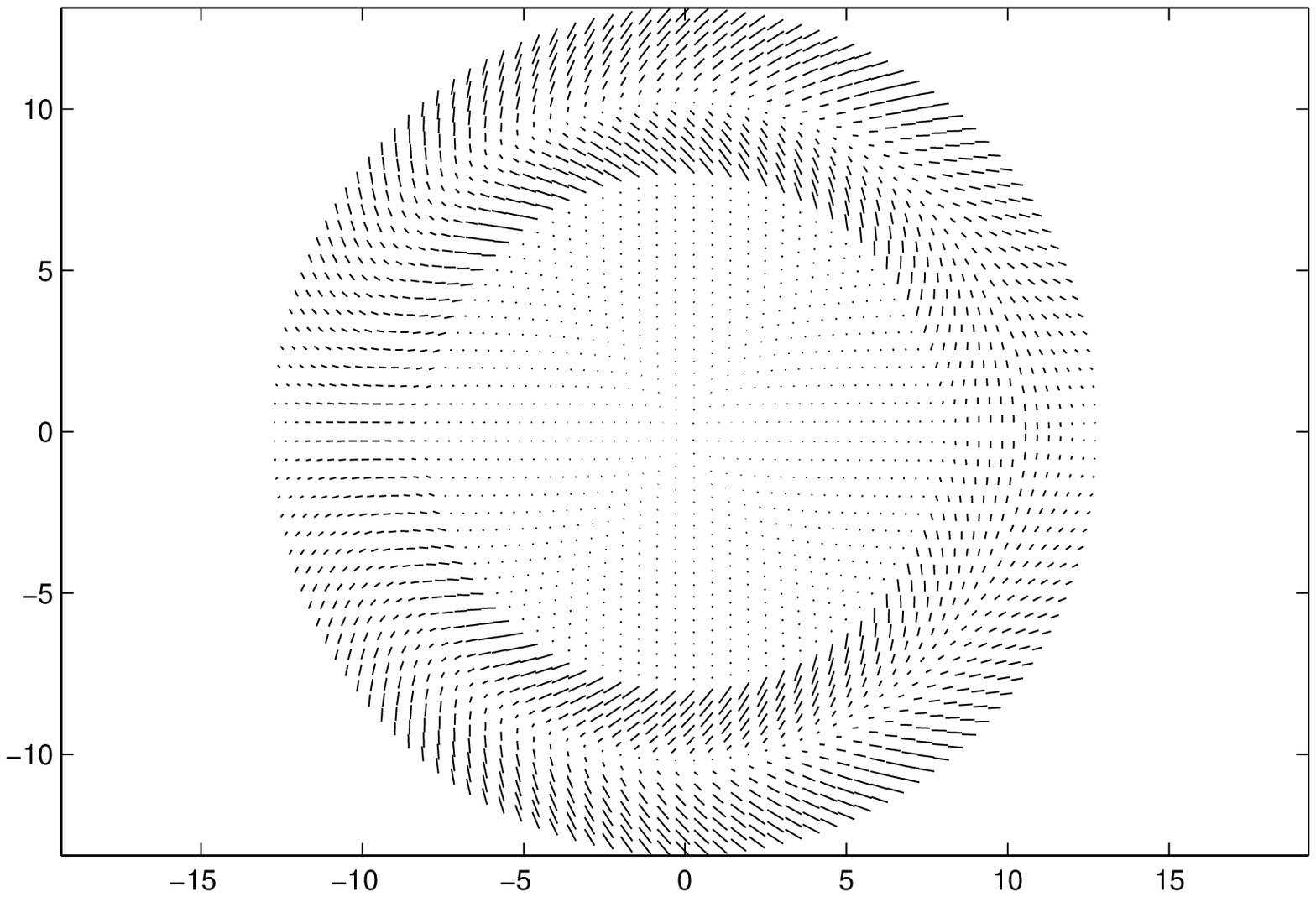} \hfill \epsfxsize=3.0truein\epsffile{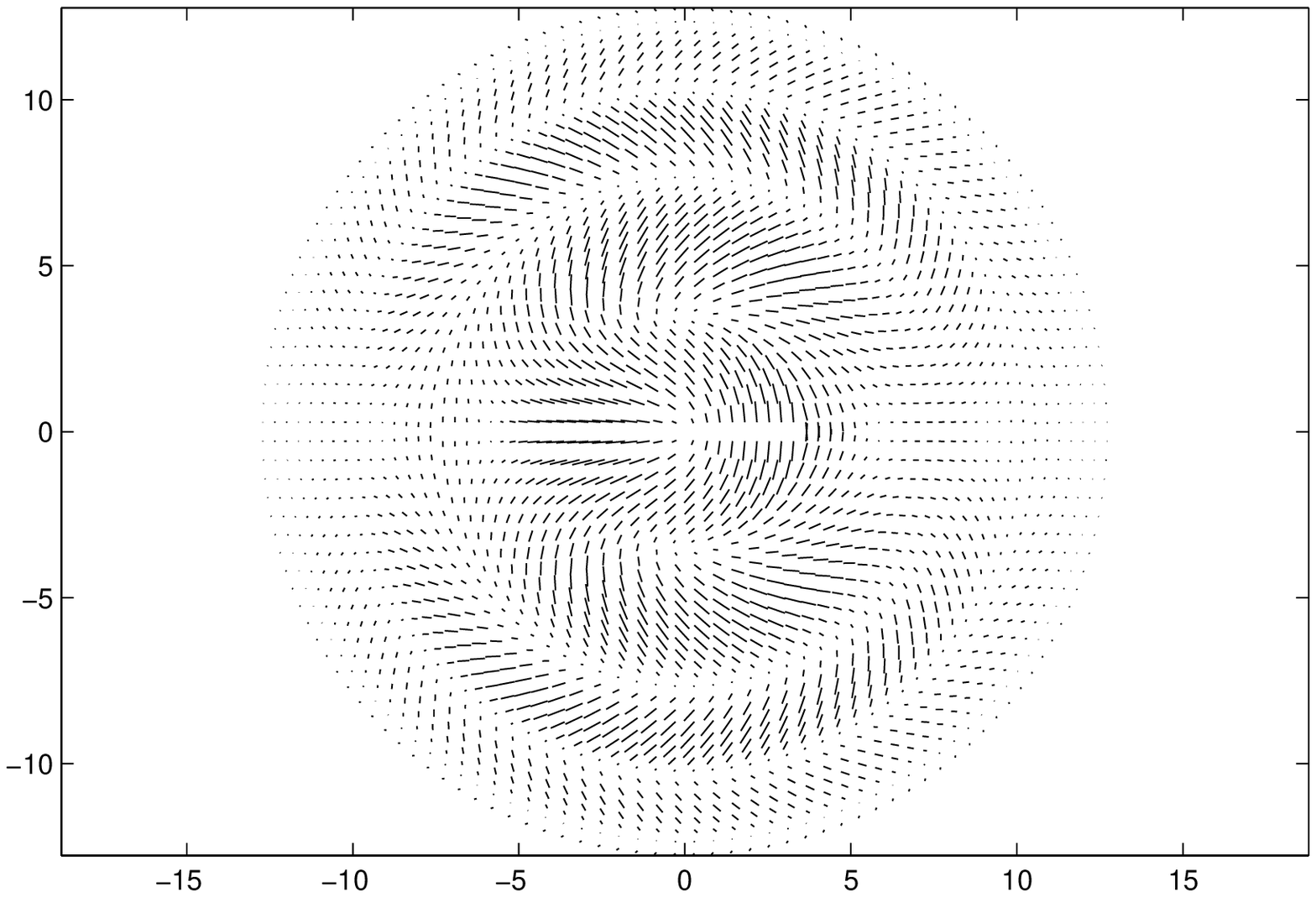} }
\caption{The pure B-mode (\ref{ex13}), for $(\ell,m) = (10,1)$ (left) and $(50,1)$ (right), using apodization length $r_* = 5^\circ$.
Only the real part is shown.  In the left panel, the second and third terms, which are concentrated in the apodization region,
dominate the main term.}
\label{fx11}
\end{figure*}

We note that the ``pure'' multipole $\widetilde B^{pure}_{\ell m}$ is defined (Eq.\ \ref{ec4}) 
as the overlap integral between $\Pi^{ab}(x)$ and the pure B-mode
\bea
&& W(x) Y^{B}_{(\ell m)ab}(x) + N'_\ell T_{ab}{}^{cd} W_c(x) Y^{G}_{(\ell m)d}(x) \label{ex13} \\
&& \qthree  + N_\ell T_{ab}{}^{cd} W_{cd}(x) Y_{\ell m}(x).  \nn
\eea
It is illuminating to consider the behavior of this mode for varying $\ell$ and $r_*$.
In Figure \ref{fx11}, the mode is shown for $(\ell,m)=(10,1)$ and $(\ell,m)=(50,1)$, taking $r_* = 7^\circ$.
At $\ell=10$, the second and third terms in (\ref{ex13}) dominate the first, 
and the mode is concentrated in the apodization region $(r-r_*)\le\theta\le r$,
which is undesirable from a signal-to-noise perspective.
At $\ell=50$, the contribution of these terms is comparable to the main term,
and the statistical weight is distributed throughout the survey region.

This illustrates a general point: the ``pure'' estimators defined in \S\ref{sc} {\em require} apodization near the survey boundary.
In the tophat limit $r_*\rightarrow 0$, the derivatives of $W(x)$ increase in magnitude; the third term in the mode (\ref{ex13}) dominates
the others, and is concentrated at the boundary.
In this limit, $\widetilde B_{\ell m}^{pure}$ formally receives no contributions from E-modes, but achieves this using counterterms which are
line integrals around the boundary, to cancel the E-B mixing.  (Compare Eq.\ (4) of \citet{LewisSep}.)
Naturally, for noisy data, this ruins the performance of the power spectrum estimator.
In a finite apodization, the counterterms in (\ref{ex13}) are effectively ``smeared'' over a nonzero area near the boundary, and the
resulting power spectrum estimators are sensible in the presence of noise.
By comparison, unmodified pseudo-$C_\ell$ estimators behave reasonably even for a tophat weight function; apodization serves the milder
purpose of reducing Fourier ringing.

Another general feature of our method is that more apodization is optimal at low $\ell$.
At $\ell=10$, Figure \ref{fx11} shows that the counterterms in (\ref{ex13}) are dominant, if $r_* = 5^\circ$.
Since increasing $r_*$ decreases the magnitude of the counterterms, this suggests that the optimal apodization length at $\ell=10$
is greater than $r_* = 5^\circ$.
In contrast, at $\ell=50$, Figure \ref{fx11} suggests that $r_* \sim 5^\circ$ is roughly optimal.

\begin{figure*}
\centerline{\epsfxsize=7.0truein\epsffile[18 144 592 430]{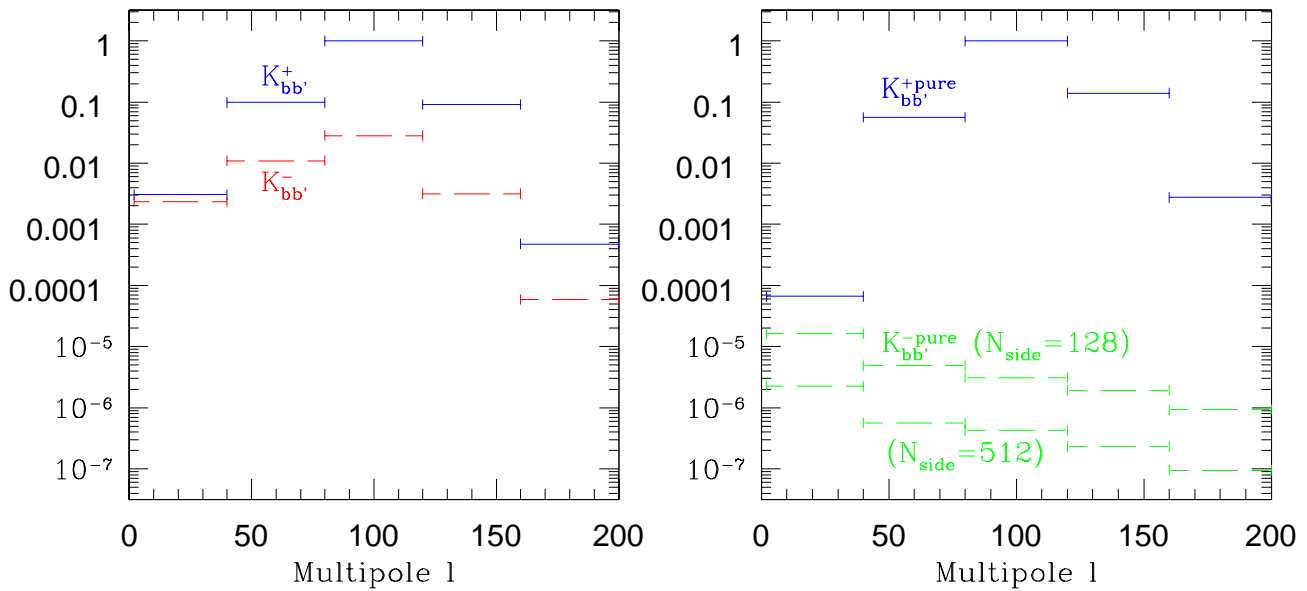}}
\caption{Transfer functions $K^\pm_{bb'}$ (left) and $K^{\pm pure}_{bb'}$ (right), shown for varying $b'$,
with the band $b$ fixed at $\ell_{min}=80$, $\ell_{max}=120$.  The apodization length used was $r_* = 5^\circ$.
The matrix entry $K^+_{bb'}$ represents the mean response of the BB estimator in band $b$ to BB power in band $b'$;
$K^-_{bb'}$ represents the response of the BB estimator to EE power.  As $N_{side} \rightarrow \infty$, $K^{-pure}_{bb'}$
approaches zero.}
\label{fx12}
\end{figure*}

Next we show the behavior of the transfer matrices $K^\pm_{bb'}$, $K^{\pm pure}_{bb'}$.
In Figure \ref{fx12}, the matrix entries are shown for varying $b'$, with the band $b$ fixed at $\ell_{min}=80$, $\ell_{max}=120$.
These matrix entries can be interpreted as the contribution to $\langle \widetilde C^{EE}_b \rangle$
and $\langle \widetilde C^{BB,pure}_b \rangle$ from E-mode and B-mode power in the band $b'$.
The small nonzero value of $K^{-pure}_{bb'}$ represents contamination of the B-mode estimators by
E-mode power in a finite pixelization, and goes to zero in the continuum limit $N_{side} \rightarrow \infty$.

\begin{figure*}
\centerline{\epsfxsize=7.0truein\epsffile[18 144 592 430]{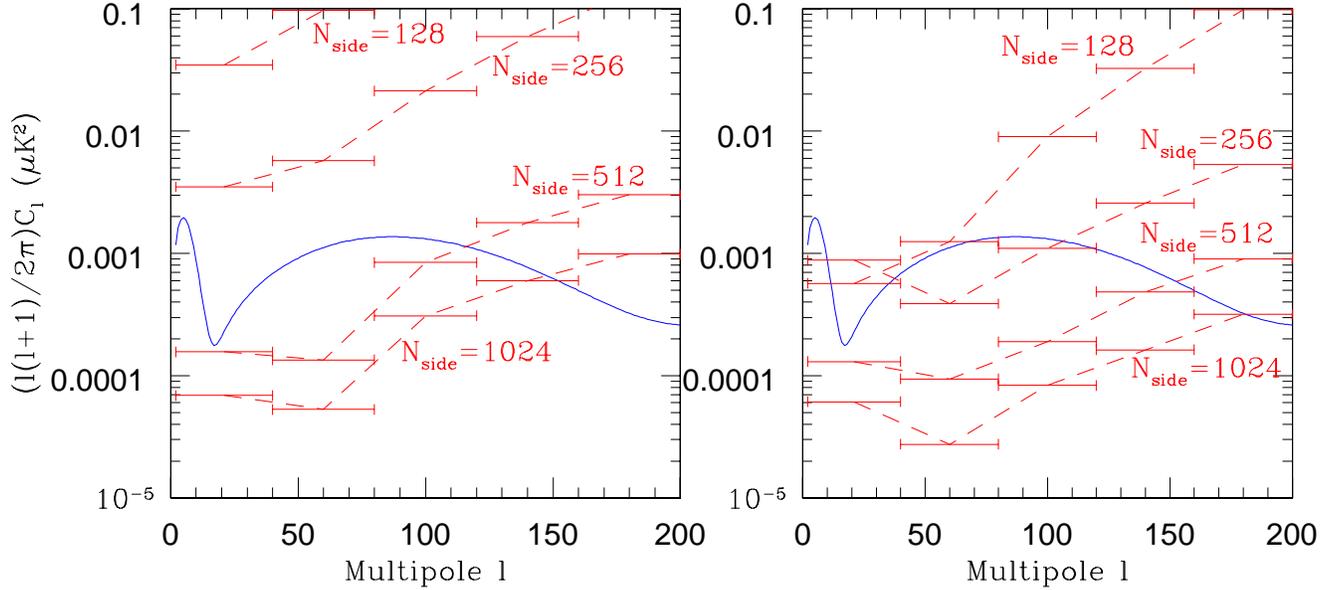}}
\caption{Contribution to $\langle \widetilde C^{BB,pure}_b \rangle$ from the fiducial E-mode power spectrum,
for zero beam (left) and FWHM 25' (right).
This can be interpreted as the estimated B-mode power which is contributed by E-modes due to pixelization artifacts.
A gravity wave B-mode spectrum with $T/S = 0.01$ is shown for scale.  The apodization length used was $r_* = 5^\circ$.}
\label{fx13}
\end{figure*}

The transfer matrix formalism can also be used to calculate the total contribution to $\langle \widetilde C^{BB,pure}_b \rangle$
from all E-mode power in the fiducial model, including power at $\ell > 200$.
One can think of this as the pseudo power spectrum of E-mode power which is aliased to B by the pixelization.
The result is shown in Figure \ref{fx13}; for comparison, a gravity wave B-mode spectrum with $T/S = 0.01$ is also shown.
As expected, the aliased power goes to zero in the continuum limit.
The level of aliased power, relative to the $T/S = 0.01$ gravity wave signal, suggests that detecting such a
small B-mode signal requires a massive amount of overpixelization.
We will quantify this better in \S\ref{sts}.

\begin{figure*}
\centerline{\epsfxsize=7.0truein\epsffile[18 144 592 430]{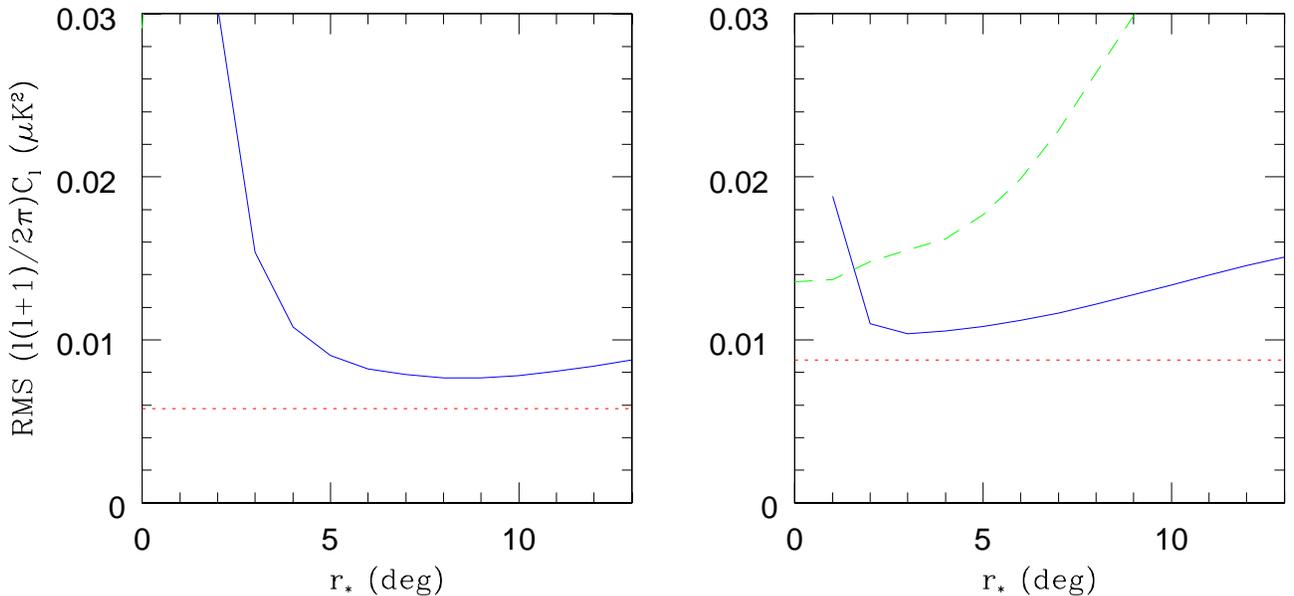}}
\caption{RMS bandpower errors for pure pseudo-$C_\ell$ estimators (blue/solid), unmodified psuedo-$C_\ell$ estimators (green/dashed),
and optimal Fisher errors (red/dotted), shown for varying apodization length $r_*$.
In the left panel, the band $10\le\ell\le 40$ is shown; in this band, the performance of unmodified pseudo-$C_\ell$ estimators is
very poor and the dashed curve is not visible.  In the right panel, the band $40 \le\ell\le 80$ is shown.
The noise level used was 20 $\mu$K-arcmin.}
\label{fx14}
\end{figure*}

We now take up the issue of choosing the apodization length $r_*$ in Eq.\ (\ref{ex12}).
In Figure \ref{fx14}, the Monte Carlo RMS scatter of the unbiased estimator $\widehat C^{BB,pure}_b$ is shown for varying $r_*$,
in two bands $b$: the lowest band $10\le\ell\le 40$, and the second lowest band $40 \le \ell \le 80$.
For both bands, the estimator performance degrades sharply when $r_*$ is chosen smaller than the optimal value,
but the optimal value is $r_* \sim 8^\circ$ for the first band and $r_* \sim 3^\circ$ for the second.
This is consistent with the qualitative discussion after Eq.\ (\ref{ex13}).
For the higher bands, repeating this analysis shows that $r_* \sim 3^\circ$ is roughly optimal.
Our solution to this problem is to use one weight function $W(x)$, with $r_* = 8^\circ$ for the lowest $\ell$ band,
and a different $W(x)$, with $r_* = 3^\circ$ for the higher bands.

We briefly describe the implementational details associated with using different weight functions in
different $\ell$ bands.  We emphasize
that these comments apply to all types of pseudo-$C_\ell$ estimators, not just the pure estimators studied in this paper.
To evaluate the estimators, one first computes a separate set of $\widetilde B_{\ell m}$ for each weight function.
Then, in each band $b$, the pseudo power spectrum $\widetilde C_b$ is computed using the $\widetilde B_{\ell m}$
corresponding to the appropriate weight function.
Generally speaking, one could use any number $N_{wt} \le N_{band}$ of weight functions, at the expense of increasing the
computational cost, since $N_{wt}$ sets of spherical harmonic transforms must be computed.
(The increase in the cost is smaller than a factor $N_{wt}$, since the transforms are $\bigoh(\ell_{max}^3)$.
For the specific example in this section, using a different weight function in the lowest band, with $\ell_{max} = 40$,
has a very small impact on the running time.)
When computing the transfer matrix element $K_{bb'}$, one uses the weight function associated to band $b$.

We have now specified the weight functions which will be used for the mock survey in this section.
To make a fair comparison between the pure and unmodified pseudo-$C_\ell$ estimators, we optimized the apodization
length $r_*$ independently for the unmodified versions, but found that a tophat weight function ($r_* = 0$) was optimal in
all bands.  
In the lowest $\ell$ band ($10\le\ell\le 40$), we found that unmodified pseudo-$C_\ell$ estimator performance was extremely poor (Figure \ref{fx14}).
If E-modes are artificially removed from the signal, then it improves dramatically, suggesting
that this is due to severe E-B mixing in the lowest band.

\begin{figure*}
\centerline{\epsfxsize=7.0truein\epsffile{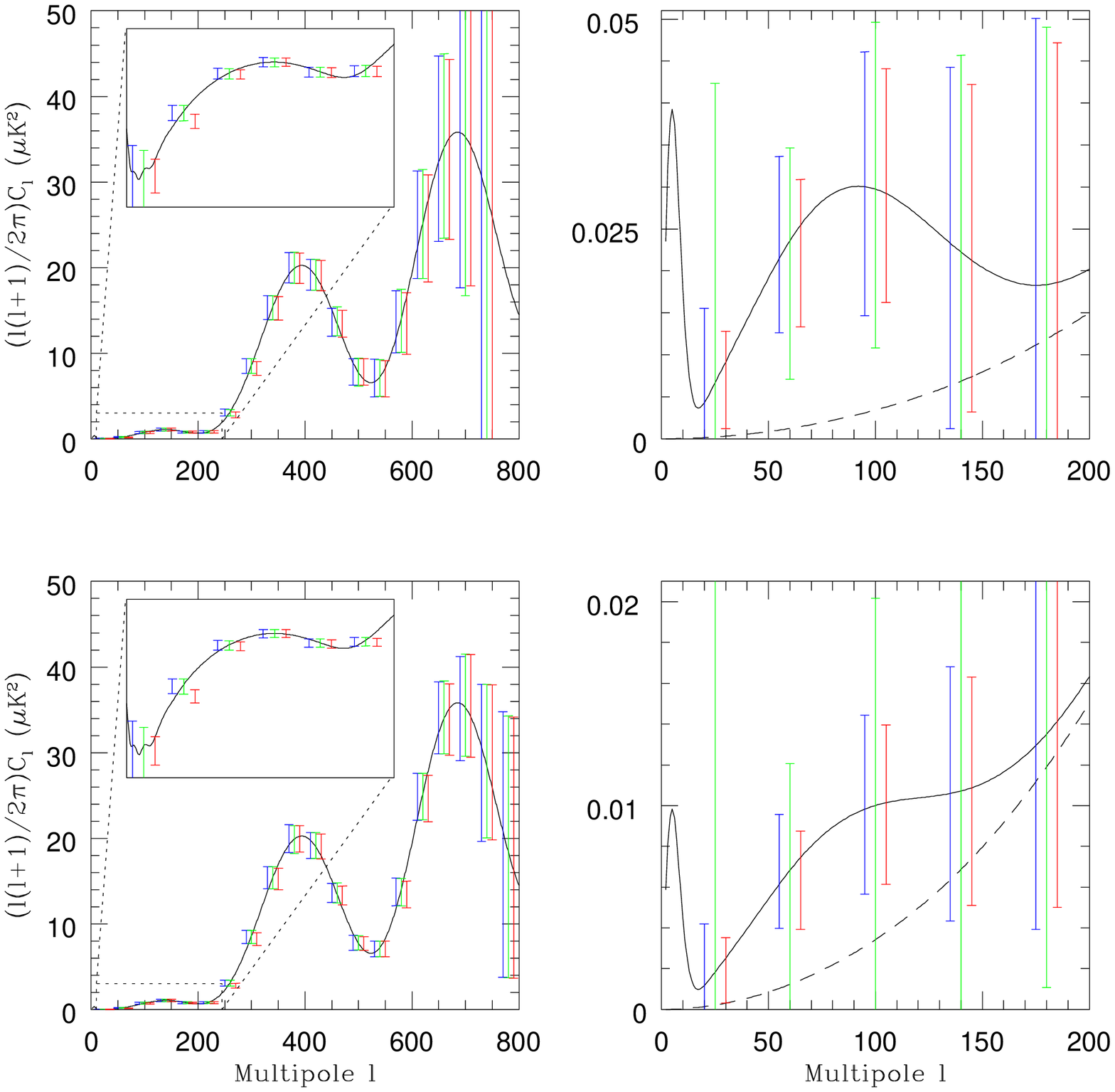}}
\caption{RMS bandpower errors for pure pseudo-$C_\ell$ estimators (blue/left), unmodified pseudo-$C_\ell$ estimators (green/middle),
and optimal Fisher errors (red/right).
The survey region is a spherical cap of radius $13^\circ$ with uniform white noise.
The top two panels are for a fiducial model with $T/S=0.2$ and noise level 20 $\mu$K-arcmin; 
the bottom two panels are for a fiducial model with
$T/S=0.05$ and noise level 10 $\mu$K-arcmin.  In each pair of panels, bandpowers are shown for E-modes (left) and B-modes (right).
In the E-mode panels, multipoles $\ell\le 250$ have been replotted on a log scale for visibility.
In the B-mode panels, the lensing component of the power spectrum has been shown separately (dashed); detecting the gravity wave signal
requires measuring power in excess of this level.}
\label{fx15}
\end{figure*}

In Figure \ref{fx15}, we have compared (Monte Carlo) RMS errors of pure pseudo-$C_\ell$ estimators, RMS errors of the
unmodified pseudo-$C_\ell$ estimators, and the Fisher bound (\ref{ex11}).
This was done for two choices of signal and noise level:
first, a $T/S=0.2$ gravity wave signal and noise level 20 $\mu$K-arcmin, 
and second, a $T/S=0.05$ gravity wave signal and noise level 10 $\mu$K-arcmin.
For E-modes, the performance of the pure and unmodified estimators is the same, as stated at the end of \S\ref{sc},
and close to optimal except in the lowest band.
For B-modes, the relative performance depends on the noise level: for the first choice,
the pure estimators perform slightly better than the unmodified versions, and slightly worse than the Fisher bound, in all bands.
The exception is the lowest band, in which the performance of the unmodified estimator is very poor.
For the second choice of noise level, the performance of the unmodified estimators has degraded significantly, but the performance of
the pure estimators remains about the same relative to optimal.

\section{An example with inhomogeneous noise}
\label{sx2}
In the previous section, we studied a mock survey with homogeneous white noise.  As a first step toward more realistic noise models,
in this section we study noise which is inhomogeneous, but not correlated between pixels.  We will use the same region as in \S\ref{sx1},
namely a spherical cap of radius $r = 13^\circ$, but the following special form for the noise covariance:
\bea
\langle Q({\bf x}) Q({\bf x'}) \rangle &=& \langle U({\bf x}) U({\bf x'}) \rangle \label{ex21}  \\
                                       &=& \eta^2 \left( \frac{r}{1-\cos r} \right) \sin(\theta)\,\delta^2({\bf x - \bf x'}) \, .       \nn
\eea
Here, $\eta$ is a constant with units $\mu$K-arcmin which we will use to quote the noise level.
The normalization $r/(1-\cos r)$ is included so that the total integrated sensitivity will be the same as a homogeneous survey whose
noise level is $\eta$.
As in \S\ref{sx1}, we will study two sensitivity levels: $\eta = 20$ $\mu$K-arcmin and $\eta = 10$ $\mu$K-arcmin.
Because the noise covariance (\ref{ex21}) is azimuthally symmetric, optimal power spectrum errors can be calculated using the 
method of Appendix \ref{afi}, and used to benchmark pseudo-$C_\ell$ power spectrum estimation.

To motivate the form (\ref{ex21}), consider a single detector which makes constant-velocity scans through the center of the
survey region, at a variety of angles.  If the distribution of angles is uniform, the scans are rapid, and the timestream noise
is white, then the resulting noise in the pixel domain will be given by (\ref{ex21}).

We must first choose pixel weight functions which will be used to construct pseudo-$C_\ell$ estimators for the noise (\ref{ex21}).
Considering E-mode estimators first, we have found that a tophat weighting is significantly suboptimal at high $\ell$, in contrast
to the homogeneous case.
(A similar result for temperature estimators appears in \cite{HGHCl}, where it is shown that Gaussian apodization improves a tophat
weight function in a mock survey with inhomogeneous noise, but not in the homogeneous case.)
However, we were able to construct E-mode estimators which were 95\% of optimal, for all $\ell \ge 80$, 
by using weight functions of the form
\be
W_\epsilon(\theta) = \frac{1}{\sin(\theta) + \epsilon}.
\label{ex22}
\ee
The form of $W_\epsilon(\theta)$ was motivated by the observation that inverse noise weighting, $W(\theta)=1/\sin(\theta)$, 
should be optimal in the noise-dominated limit; we included a regulator $\epsilon$ to smooth the singularity at $\theta=0$.
The values of $\epsilon$ used are shown in Table \ref{tx21}.
We were unable to achieve the same level of EE estimator performance using cosine apodization (\ref{ex12}) in place of the apodization (\ref{ex22}),
or using fewer than four different weight functions.

\begin{table}
\begin{center}
\begin{tabular}{|cc|}
\hline
\multicolumn{2}{|c|}{$\eta = 20$ $\mu$K-arcmin}  \\
$\ell$ range          &   weight function                          \\
\hline
$\ell\le 440$         &   tophat weighting                        \\
$400\le\ell\le 560$   &   $W_\epsilon(\theta)$, $\epsilon=0.070$   \\
$560\le\ell\le 680$   &   $W_\epsilon(\theta)$, $\epsilon=0.025$   \\
$\ell\ge 680$         &   $W_\epsilon(\theta)$, $\epsilon=0.010$   \\
\hline
\multicolumn{2}{|c|}{$\eta = 10$ $\mu$K-arcmin}  \\
\hline
$\ell\le 600$        &  tophat weighting                       \\
$600\le\ell\le 720$  &  $W_\epsilon(\theta)$, $\epsilon=0.070$  \\
$720\le\ell\le 800$  &  $W_\epsilon(\theta)$, $\epsilon=0.017$  \\
$\ell\ge 800$        &  $W_\epsilon(\theta)$, $\epsilon=0.008$  \\
\hline
\end{tabular}
\end{center}
\caption{Pixel weight functions used to construct E-mode estimators for the mock surveys in this section.}
\label{tx21}
\end{table}

This example highlights a practical issue for pseudo-$C_\ell$ power spectrum estimation: for a given survey, how does one
choose pixel weight functions which optimize performance of the estimators?
This issue is separate from the E-B separation problem which is the focus of this paper; we have seen here that it is important
even for E-mode power spectrum estimation alone.
In a future paper, we plan to investigate algorithms for constructing optimal pixel weight functions, starting from Monte Carlo
simulations of the noise.

Turning now to weight functions for B-mode estimators, we have found that cosine apodization (\ref{ex12})
results in pseudo-$C_\ell$ estimators whose performance (relative to optimal) is similar to the homogeneous case.
However, best results were obtained using apodization length $r_* = r = 13^\circ$ in all bands, in contrast to the homogeneous case
where we used $r_* = 8^\circ$ for $10\le\ell\le 40$ and $r_* = 3^\circ$ for $\ell\ge 40$.

\begin{figure*}
\centerline{\epsfxsize=7.0truein\epsffile{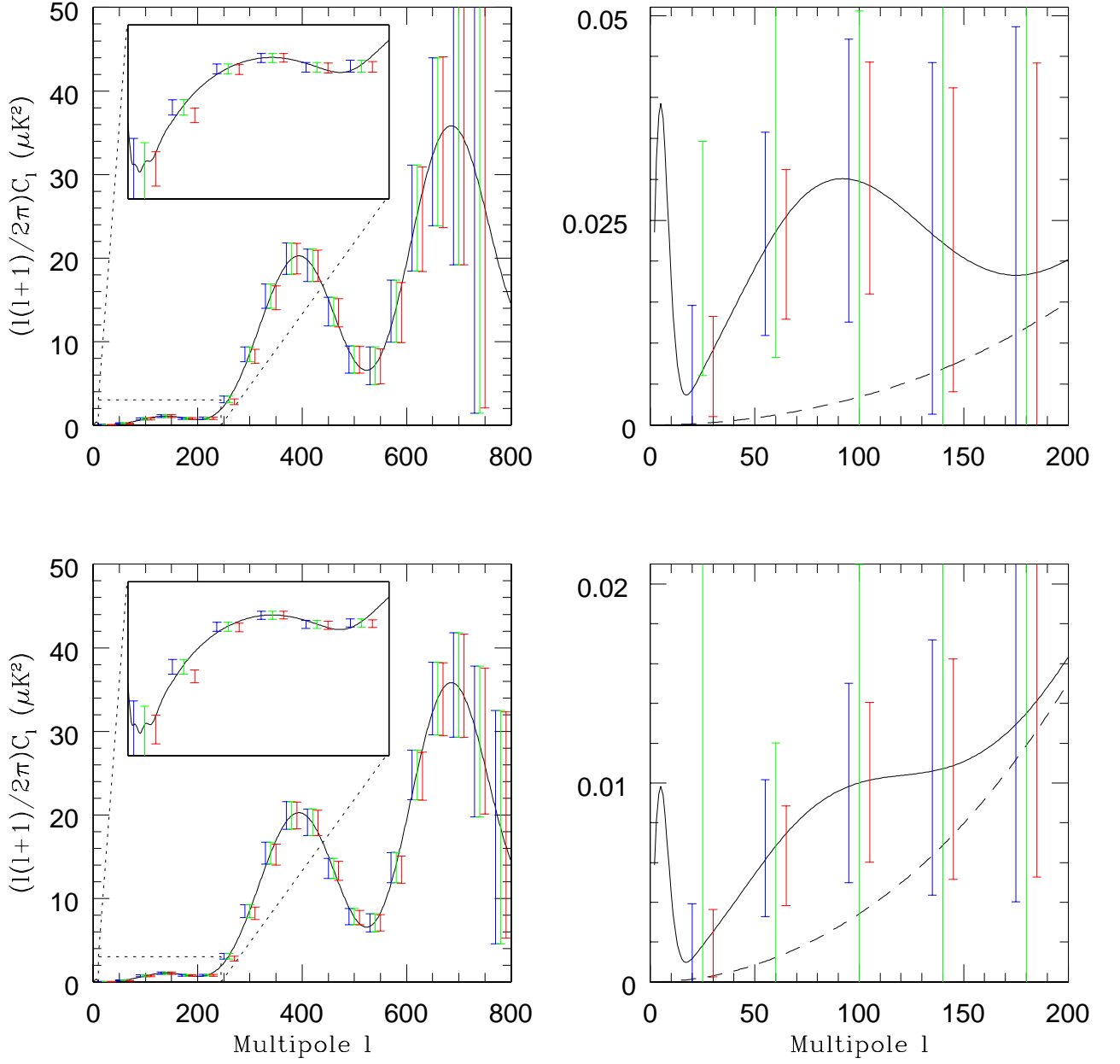}}
\caption{RMS bandpower errors for pure pseudo-$C_\ell$ estimators (blue/left), unmodified pseudo-$C_\ell$ estimators (green/middle),
and optimal Fisher errors (red/right).
The survey region is a spherical cap of radius $13^\circ$ with inhomogeneous noise given by (\ref{ex21}).
The top two panels are for a fiducial model with $T/S=0.2$ and noise level 20 $\mu$K-arcmin; 
the bottom two panels are for a fiducial model with $T/S=0.05$ and noise level 10 $\mu$K-arcmin.  
In each pair of panels, bandpowers are shown for E-modes (left) and B-modes (right).}
\label{fx21}
\end{figure*}

We have now completely specified weight functions; the performance of the pseudo-$C_\ell$ estimators is shown in Figure \ref{fx21}.
The results are similar to the homogeneous case.
At sensitivity $\eta = 20$ $\mu$K-arcmin, the pure B-mode estimators perform slightly better than the unmodified estimators, and slightly
worse then optimal.  
At increased sensitivity $\eta = 10$ $\mu$K-arcmin, the performance of the pure B-mode estimators remains the same relative to optimal, but
the unmodified estimators have significantly degraded.
Our conclusion in this section is that inhomogeneous noise, at least for the specific form (\ref{ex21}) studied here, does not pose a
problem in principle for power spectrum estimation using pure pseudo-$C_\ell$ estimators, 
although the practical issue of choosing suitable pixel weight functions remains.

\section{Detectability of gravity wave B-modes}
\label{sts}

\begin{figure*}
\centerline{\epsfxsize=7.0truein\epsffile[18 144 592 430]{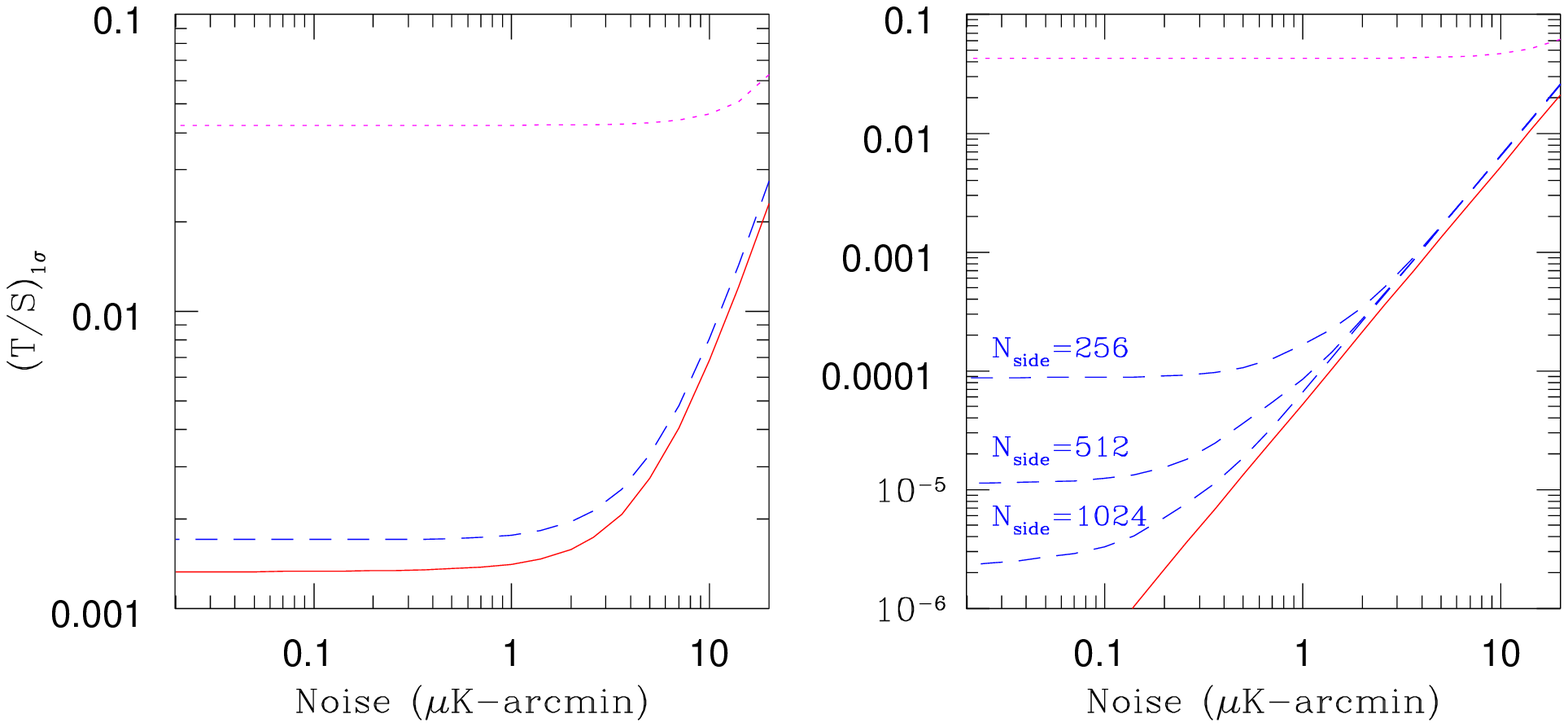}}
\caption{Minimum T/S detectable at $1\sigma$, for a $13^\circ$ spherical cap survey with uniform noise, assuming a 25' beam
and Gaussian lensing contaminant (left), and zero beam and no lensing contamination (right).  The solid line is the optimal
Fisher value, the dotted line is the Monte Carlo value using unmodified pseudo-$C_\ell$ estimators (\S\ref{se}), and the dashed lines
are Monte Carlo values using pure pseudo-$C_\ell$ estimators.
For the parameters in the right panel, finite-pixelization artifacts impose a ``floor'' on the value of T/S which can be
detected, but the value can be made arbitrarily small by increasing the resolution.
For the parameters in the left panel, we find that increasing the resolution beyond $N_{side} = 256$ results in no significant
improvement.}
\label{fts1}
\end{figure*}

\par\noindent
In this section, we will study the smallest gravity wave B-mode signal which can be detected using our pure pseudo-$C_\ell$
estimators, as a function of the noise level.
We use the survey parameters of \S\ref{sx1}:  a $13^\circ$ spherical cap
with uniform noise, and define pseudo-$C_\ell$ estimators using the $\ell$ bands and pixel weight
functions defined in that section.

In the left panel of Figure \ref{fts1}, we have considered a ``wide-beam'' survey with a 25' Gaussian beam.
The lensing B-mode signal is treated as an extra source of Gaussian noise.
The minimum $(T/S)$ which is detectable at $1\sigma$ is given by
\be
(T/S)_{1\sigma} = \left[ (\delta E)^{b} C^{-1}_{bb'} (\delta E)^{b'} \right]^{-1/2},
\ee
where $C^{bb'}$ denotes the bandpower covariance of the estimators (including noise contribution) in a fiducial model with $T/S=0$,
and $(\delta E)^b$ denotes the estimator mean contributed by a tensor $B$-mode signal with $T/S=1$.
We note that $(\delta E)^b$ can be computed exactly using the transfer matrix formalism of Appendix \ref{atr},
but $C^{bb'}$ must be computed by Monte Carlo.

In Figure \ref{fts1} (left panel), we show the minimum detectable $T/S$ for both pure and unmodified pseudo-$C_\ell$
estimators, with the optimal Fisher value (computed using the method of Appendix \ref{afi}) shown for comparison.
In all three cases, there is a floor to the gravity wave signal which can be detected, arising from lensing contamination,
even in the limit of zero noise.
With unmodified pseudo-$C_\ell$ estimators, we find the best possible $1\sigma$ detection is $T/S = 0.042$, which is reached
at a total sensitivity of $\sim 10$ $\mu$K-arcmin.  This agrees well with the results of \citet{ChalChon}.
With pure psuedo-$C_\ell$ estimators, the best possible $1\sigma$ detection is $T/S = 0.00170$, which is reached at total sensitivity 
$\sim 2$ $\mu$K-arcmin.  For comparison, the optimal value is $T/S = 0.00133$, or $78\%$, so the performance of the estimators is close to
optimal.
These $(T/S)$ values are specific to the survey region considered here, and depend on the geometry in a way which involves
boundary E-B mixing and does not scale as $f_{sky}^{-1/2}$ as mode-counting arguments would suggest \citep{AHS}.

In principle, these values of $(T/S)$ are not necessarily ultimate limits, since we have arrived at them by treating the
lensing B-modes as a Gaussian contaminant which can not be separated from the gravity wave signal.
In actuality, lensing B-modes are non-Gaussian, and ``delensing'' algorithms have been proposed \citep{OHdelens,HSdelens}
which exploit this to separate the gravity wave and lensing components, reducing the level of contamination.
The amount of lensing contamination which can be removed in this way depends on the noise level and beam size.
In this paper, we make no attempt to model residual delensing errors realistically.
Instead, we note that the result of delensing will lie between two extremes: no separation of the gravity wave and
lensing B-modes (which has just been considered), and complete separation.
In the right panel of Figure \ref{fts1}, we consider the latter extreme; we completely remove the lensing component
of the B-mode power spectrum, and show $(T/S)_{1\sigma}$ as a function of noise level.
We have also used zero beam size, to crudely reflect the fact that delensing algorithms require measuring modes at high $\ell$
to separate the gravity wave and lensing signals at low $\ell$.
Under these extreme assumptions, there is no limit
in our framework to the detectability of gravity wave B-modes as the signal-to-noise improves, since we do not consider
other sources of contamination such as astrophysical foregrounds.

Using unmodified pseudo-$C_\ell$ estimators, the smallest gravity wave signal which can be detected is
the same as in the left panel, $T/S = 0.042$; removing the lensing component of the B-modes results in no improvement.
Rephrasing, the extra BB estimator covariance which is contributed by E-modes dominates the contribution from lensing B-modes.
Using pure pseudo-$C_\ell$ estimators, there is a floor to $(T/S)_{1\sigma}$ in any
fixed pixelization, but the floor improves without limit as the resolution is increased ($N_{side} = 256, 512, 1024$ are shown.)
This is consistent with the discussion in \S\ref{sx1}; in a finite pixelization there is some contamination of the B-mode
estimators by E-modes, but the contamination goes to zero in the continuum limit.
For any fixed noise level, one can choose a sufficiently high resolution so that the contamination is negligible.
If this is done, then the value of $(T/S)_{1\sigma}$ obtained is approximately 80\% of the optimal Fisher value, for all noise
levels considered in the right panel of Figure \ref{fts1}.

The results of this section illustrate a qualitative difference between our pure pseudo-$C_\ell$ estimators
and the unmodified versions.  As the signal-to-noise improves in a fixed survey region, a floor is revealed
to the gravity wave signal which can be detected using unmodified pseudo-$C_\ell$ estimators.
This is not the case for pure pseudo-$C_\ell$'s, although a second smaller floor is eventually revealed 
when the signal-to-noise becomes good enough that the sensitivity is limited by contamination from lensing.
Going beyond this will require separating the lensing signal using delensing algorithms.
We have crudely studied this regime by assuming perfect delensing and zero beam size; under these assumptions,
we find that the estimators alone impose no limit to the value of $(T/S)$ which can be detected.

\section{Conclusion}
\label{sco}

In this paper, we have defined pure pseudo-$C_\ell$ estimators, which have the property that the
estimated B-mode power receives no contribution from E-modes, even on a cut sky.
As a consequence, E-mode signal power does not contribute to the variance of the B-mode estimators.
The pixel weight functions for these estimators have spin-1 and spin-2 components (\ref{ec5}),
which can be obtained from the spin-0 component by covariant differentiation if its form is analytical, 
or finite differencing (Appendix \ref{afd}) in general.
In contrast to the usual pseudo-$C_\ell$ formalism, the spin-0 weight function must have a finite apodization
near the boundary, and obey Dirichlet and Neumann boundary conditions.
We give an algorithm (Appendix \ref{atr}) for computing the pseudo-$C_\ell$ transfer matrix which is used for
debiasing, and show that its computational cost is $\bigoh(\ell_{max}^3)$, which is the same as the cost of 
evaluating the estimators.

We have studied these estimators in detail for mock surveys on a $13^\circ$ spherical cap, using both
homogeneous noise, and inhomogeneous noise of a specific form (\ref{ex21}).
In both cases, we found that for B-mode power spectrum estimation,
pure pseudo-$C_\ell$ estimators performed slightly better than the unmodified versions at noise level 20 $\mu$K-arcmin,
and much better at noise level 10 $\mu$K-arcmin.
In the homogeneous case, we considered a wide range of noise levels, and showed that the pure estimators are
$\sim 80\%$ of optimal, defined by the degradation in $(T/S)_{1\sigma}$ which can be detected.  The $80\%$
figure is obtained in two limiting cases: assuming no separation of the gravity wave and lensing signals,
and perfect separation.  In the latter case, there is no limit, imposed by the estimators alone,
to the value of $(T/S)$ which can be detected.
In constrast, using unmodified pseudo-$C_\ell$ estimators, we have found that the smallest signal which can be detected
in a patch with this geometry is $(T/S)_{1\sigma} = 0.042$, in agreement with \citet{ChalChon}.

In future work, we plan to investigate two basic issues left unaddressed by this paper.
First, we will study the prospect of generating optimal weight functions for a survey algorithmically, starting from Monte 
Carlo simulations of the noise, rather than choosing them heuristically.
The need for such a procedure is highlighted by the inhomogeneous noise model considered in \S\ref{sx2},
in which we were forced to use four weight functions with different levels of apodization, in order to
achieve near-optimal E-mode estimation in all $\ell$ bands.
Second, we will study pure pseudo-$C_\ell$ estimators for more realistic noise models, including models
which include correlated noise.

\section*{Acknowledgments}
We would like to thank Wayne Hu and Bruce Winstein for many discussions throughout this project.
We also thank the referree for careful proofreading of the manuscript and helpful suggestions.
We acknowledge use of the FFTW, LAPACK, CAMB, and Healpix software packages.
This work was supported by the Kavli Institure for Cosmological Physics through the grant NSF PHY-0114422.

\appendix

\section{Spin-1 spherical harmonic transforms}
\label{avh}

As discussed at the end of \S\ref{sc2}, our algorithm for fast evaluation of $\widetilde B_{\ell m}$
requires an implementation of fast spherical harmonic transforms for vector fields.
This will also be an ingredient in the algorithm, to be presented in Appendix \ref{atr},
for computing the transfer matrices $K^{\pm pure}_{\ell\ell'}$.
In this appendix, we write down the recursion and initial conditions which are needed:
\bea
({}_1\lambda_{\ell+1,m}) Y^G_{\ell+1,m} &=& (\cos\theta)Y^G_{\ell m} - \frac{im}{\ell(\ell+1)}Y^C_{\ell m}   \\
                                         && \qtwo - ({}_1\lambda_{\ell m}) Y^G_{\ell-1,m}                  \nn \\
({}_1\lambda_{\ell+1,m}) Y^C_{\ell+1,m} &=& (\cos\theta)Y^C_{\ell m} + \frac{im}{\ell(\ell+1)}Y^G_{\ell m}   \nn \\
                                         && \qtwo - ({}_1\lambda_{\ell m}) Y^C_{\ell-1,m}                  \nn \\
(Y^G_{\ell\ell})_a = \epsilon_{ab} (Y^C_{\ell\ell})^b &=& \frac{(-1)^\ell}{2^\ell \ell!} \sqrt{\frac{(2\ell+1)!\ell}{4\pi(\ell+1)}}  \times \nn \\
                                         && \qquad (\sin^{\ell-1}\theta) e^{i\ell\varphi} \big[ (\cos\theta) x_a + i y_a \big]        \nn
\eea
Here, $({}_1\lambda_{\ell m}) = (1/\ell) \sqrt{(\ell^2-1)(\ell^2-m^2)/(4\ell^2-1)}$.

An alternate approach to implementing fast spin-1 transforms appears in \citep{LewisLensing}.

\section{Finite differencing in an irregular spherical pixelization}
\label{afd}

Our definition of $\widetilde B_{\ell m}$ (\ref{ec4}) requires a prescription for computing the spin-1 and spin-2 weights $W_a$, $W_{ab}$
(\ref{ec5}) from $W(x)$.  If $W(x)$ is of known analytical form, then a trivial ``prescription'' consists of carrying out the covariant
derivatives by hand; otherwise, a finite differencing scheme must be used.  Since this is nontrivial in an irregular spherical pixelization
such as Healpix, we present the details of one such scheme here.

At each pixel center $\bx$, we define a matrix $M_{ab}(\bx)$ by
\be
M^{ab}(\bx) = \sum_{\bx'} (\bx'-\bx)^{\perp a} (\bx'-\bx)^{\perp b}
\ee
Throughout this appendix, the notation $\sum_{\bx'}$ denotes a sum over pixel neighbors $\bx'$ to the pixel $\bx$,
and $(\bx'-\bx)^\perp$ denotes projection of the 3-vector $(\bx'-\bx)$ into the plane
perpendicular to the unit vector $\bx$.  After the projection, $(\bx'-\bx)^\perp$ is a tangent vector at $\bx$, and $M$ is
a rank-2 symmetric tensor, as the index notation suggests.

Now, if $W(\bx)$ is any pixelized map, we define its ``finite difference'' gradient by
\bea
\nabla^{FD}_a W(\bx) &\eqdef& M_{ab}^{-1}(\bx) \times  \\
&&\qquad \left[ \sum_{\bx'} (f(\bx')-f(\bx)) (\bx'-\bx)^{\perp b} \right]  \nn
\eea
For purposes of this paper, we need the following extension.  If $v_a(\bx)$ is a finite pixelized (tangent) vector field, we define
its ``finite difference'' covariant derivative by
\bea
\nabla^{FD}_a v_b(\bx) &\eqdef& M_{ac}^{-1}(\bx)  \times \\
&& \quad \left[ \sum_{\bx'} ({\mathcal P}_{\bx'\rightarrow\bx}v(\bx')-v(\bx))_b (\bx'-\bx)^{\perp c} \right]  \nn
\eea
where ${\mathcal P}_{\bx'\rightarrow\bx} v(\bx')$ denotes the tangent vector at $\bx$ obtained by parallel translating $v(\bx')$ along the
great circle arc connecting $\bx$ and $\bx'$.  This parallel translation is needed to covariantly difference tangent vectors at
distinct points $\bx$, $\bx'$.  If one were to simply difference their components in the $\theta$, $\varphi$ coordinate system instead, then the
resulting finite differencing scheme would suffer from coordinate artifacts near the poles.

Using this finite differencing scheme, one takes the spin-1 weight $W_a$ in Eq.\ (\ref{ec5}) to be 
$\nabla^{FD}_a W$, and the spin-2 weight $W_{ab}$ to be the traceless symmetric part of $\nabla^{FD}_a (\nabla^{FD}_b W)$.  
We have tested this prescription for the case of cosine apodization, by comparing the pure pseudo-$C_\ell$ estimators 
which result from finite differencing and analytic differentiation of the cosine weight function.
For a fixed random CMB realization, we find that the difference between the two versions is negligible.

\section{Correlation functions for fields of arbitrary spin}
\label{asp}
For the transfer matrix calculations in Appendix \ref{atr}, we need expressions for correlation functions
in terms of power spectra, for all combinations of spin-0, spin-1, and spin-2 
fields.
We present these results in a form which generalizes to higher spins as well.
A spin $s$ field ($-\infty < s < \infty$) is a function $({}_sf)$
whose value at $x$ depends on a choice of orthonormal basis vectors
$\{ \e_1, \e_2 \}$ at $x$.  Under the right-handed rotation
\begin{eqnarray}
\e'_1 &=&  (\cos\theta)\e_1 + (\sin\theta)\e_2    \\
\e'_2 &=& -(\sin\theta)\e_1 + (\cos\theta)\e_2   \nonumber
\end{eqnarray}
$({}_sf)$ must transform as $({}_sf)' = e^{-is\theta} ({}_sf)$.
There is a spin-raising operator $\sraise$ and a spin-lowering operator $\slower$ which
transform a spin $s$ field into fields of spin $(s+1)$ and $(s-1)$ respectively.
In the frame $\{ \e_1, \e_2 \} = \{ x_a, y_a \}$, these operators take the form
\begin{eqnarray}
\sraise({}_sf) &=& -\sin^s(\theta) \left( \frac{\partial}{\partial\theta} 
                                      + i \csc(\theta) \frac{\partial}{\partial\varphi} \right) \sin^{-s}(\theta) ({}_sf)   \nn \\
\slower({}_sf) &=& -\sin^{-s}(\theta) \left( \frac{\partial}{\partial\theta} 
                                      - i \csc(\theta) \frac{\partial}{\partial\varphi} \right) \sin^s(\theta) ({}_sf)  \nn
\end{eqnarray}
An orthonormal basis for spin $s$ fields is given by the spin harmonics ${}_sY_{\ell m}$ \citep{NPspins,Goldberg}, which are given for $s > 0$ by
\bea
{}_sY_{\ell m}    &=& \sqrt{\frac{(\ell-s)!}{(\ell+s)!}} (\sraise)^s (Y_{\ell m})          \\
{}_{-s}Y_{\ell m} &=& (-1)^s \sqrt{\frac{(\ell-s)!}{(\ell+s)!}} (\slower)^s (Y_{\ell m})   \nn
\eea
These are nonzero only for $\ell \ge |s|$.

Correlation functions between fields of spin $s$, $s'$ can be expressed in terms of the following
function, which generalizes the Legendre polynomial $P_\ell(\bx\cdot \bx')$ in the case $s = s' = 0$.
\bea
P_\ell^{ss'}(\bx\cdot \bx') &\eqdef& \frac{4\pi}{2\ell+1} \sum_{m=-\ell}^\ell
                     \big( {}_{s}Y_{\ell m}(\bx)_{\{\e_1,\e_2\}=\{X,Y\}} \big)^*   \nn  \\
               &&  \qquad \times \big( {}_{s'}Y_{\ell m}(\bx')_{\{\e_1,\e_2\}=\{X',Y'\}} \big)   \label{esp6}
\eea
We emphasize that the spin harmonics on the right-hand side are evaluated in the ``two-point'' frame
$\{\e_1,\e_2\} = \{X,Y\}$, $\{\e'_1,\e'_2\} = \{X',Y'\}$, not the frame $\{\e_1,\e_2\}=\{x_a,y_a\}$.
With this choice of frame, the right-hand side is rotationally invariant and therefore depends only on the
separation $(\bx\cdot \bx')$, as implied on the left-hand side.

It can be shown \cite[Eq.~(3.17)]{NgLiu} that $P_\ell^{ss'}$ is related to the reduced Wigner D-function $d_{ss'}^\ell$ by
$P_\ell^{ss'}(\cos \theta) = (-1)^s d_{ss'}^\ell(\theta)$.
Using standard results on Wigner D-functions \cite{QTAM}, one obtains the spin-label symmetries, orthogonality relation,
and product rule:
\bea
P_\ell^{ss'}(z) = P_\ell^{s's}(z) &=& (-1)^{l+s} P_\ell^{s,-s'}(-z)  \label{esp2}  \\
\int_{-1}^1 dz\, P_{\ell_1}^{ss'}(z) P_{\ell_2}^{ss'}(z) &=& \frac{2}{2\ell_1+1} \delta_{\ell_1\ell_2}   \label{esp14}
\eea
\bea
&&  P^{s_1s_1'}_{\ell_1}(z) P^{s_2s_2'}_{\ell_2}(z) = (-1)^{s_1+s_1'+s_2+s_2'} \times                  \label{esp13}   \\
&&  \qquad \sum_{\ell_3} (2\ell_3+1) \threej{\ell_1}{\ell_2}{\ell_3}{-s_1}{-s_2}{s_1+s_2} \times                  \nn  \\
&&  \qthree \threej{\ell_1}{\ell_2}{\ell_3}{-s_1'}{-s_2'}{s_1'+s_2'} P^{s_1+s_2,s_1'+s_2'}_{\ell_3}(z)          \nn
\eea
Likewise, the following recursion and initial conditions may be used to evaluate $P^{ss'}_\ell$.
Using the spin-label symmetries (\ref{esp2}), it suffices to give initial conditions 
for the recusion assuming $s \ge |s'|$.
\bea
\rho^{ss'}_{\ell+1} P^{ss'}_{\ell+1}(z) &=& (2\ell + 1) \left[ z - \frac{ss'}{\ell(\ell+1)} \right] P^{ss'}_\ell(z) \label{esp1}  \\
&& \qtwo  - \rho^{ss'}_\ell P^{ss'}_{\ell-1}(z)    \nn  \\
P^{ss'}_s(z) &=& \frac{(-1)^{s'}}{2^s} \binom{2s}{s+s'}^{1/2} \!\! (1+z)^{(s+s')/2}   \label{esp3}  \\
&& \qquad \times (1-z)^{(s-s')/2} \qquad (s \ge |s'|)    \nn
\eea
In the recursion~(\ref{esp1}), we have defined $\rho^{ss'}_\ell = \sqrt{ (\ell^2-s^2) (\ell^2-s'^2) } / \ell$.

The recursion (\ref{esp1}) with accompanying initial condition (\ref{esp3}) is the main result of this appendix.
We have used the language of spin-$s$ fields, which permits the result to be stated in a uniform way for all spins.
Since the transfer matrix calculations of Appendix \ref{atr} use the (equivalent) tensor language in spins $\le 2$,
the rest of the appendix is devoted to translating between the two.
As a byproduct, we will obtain expressions for the correlation functions between all combinations of
fields with spins $s,s' \le 2$.
In the case where both spins are nonzero, it will be convenient to use the linear combinations
\bea
Q^{ss'}_\ell &=& \frac{P^{ss'}_\ell + (-1)^{s'}P^{s,-s'}_\ell}{2}  \label{esp15}  \\
R^{ss'}_\ell &=& \frac{P^{ss'}_\ell - (-1)^{s'}P^{s,-s'}_\ell}{2}  \nn
\eea
instead of $P^{ss'}_\ell$, $P^{s,-s'}_\ell$.

The basic observation which relates the tensor and spin-$s$ formalisms is that the frame-dependent vectors
\be
{\bf m} = (\e_1 + i\e_2)/2  \qquad\qquad \bar{\bf m} = (\e_1 - i\e_2)/2
\ee
have spins 1 and -1 respectively.
(This notation follows \citet{OHdelens} but our normalization differs by a factor $1/\sqrt{2}$.)
In terms of these, the vector and tensor harmonics $Y^G_{\ell m}$, $Y^E_{\ell m}$ can be written
\begin{eqnarray}
Y^G_{(\ell m)a}  &=& ({}_{-1}Y_{\ell m}){\bf m}_a  - ({}_1Y_{\ell m})\bar{\bf m}_a  \label{esp4}     \\
Y^E_{(\ell m)bc} &=& -({}_{-2}Y_{\ell m}){\bf m}_b{\bf m}_c - ({}_2Y_{\ell m})\bar{\bf m}_b\bar{\bf m}_c.    \nonumber
\end{eqnarray}
These are parity-even combinations of $\{ {\bf m}_a, {\bar{\bf m}}_a \}$ and 
$\{ {\bf m}_b{\bf m}_c, \bar{\bf m}_b \bar{\bf m}_c \}$.

The following sums can be evaluated using (\ref{esp4}) and the definition
(\ref{esp6}) of $P^{ss'}_\ell$:
\bea
&& \sum_m Y^{*}_{\ell m}(\bx)      Y_{\ell m}(\bx')         =  \frac{2\ell+1}{4\pi} P_\ell^{00}(z)   \label{esp7}    \\
&& \sum_m Y^{*}_{\ell m}(\bx)      Y^G_{(\ell m)a'}(\bx')   = -\frac{2\ell+1}{4\pi} P_\ell^{01}(z) X'_{a'}       \nn \\
&& \sum_m Y^{*}_{\ell m}(\bx)      Y^E_{(\ell m)b'c'}(\bx') = -\frac{2\ell+1}{4\pi} P_\ell^{02}(z) Q'_{b'c'}     \nn \\
&&           \sum_m Y^{G*}_{(\ell m)a}(\bx)  Y^G_{(\ell m)a'}(\bx')  = \frac{2\ell+1}{4\pi} \times       \nn  \\
&& \qthree          (Q^{11}_\ell(z) X_a X'_{a'} + R^{11}_\ell(z) Y_a Y'_{a'})                            \nn  \\
&&           \sum_m Y^{G*}_{(\ell m)a}(\bx)  Y^E_{(\ell m)b'c'}(\bx') = \frac{2\ell+1}{4\pi} \times      \nn  \\
&& \qthree          (Q^{12}_\ell(z) X_a Q'_{b'c'} + R^{12}_\ell(z) Y_a U'_{b'c'})                        \nn  \\
&&           \sum_m Y^{E*}_{(\ell m)bc}(\bx) Y^E_{(\ell m)b'c'}(\bx')  = \frac{2\ell+1}{4\pi} \times     \nn  \\
&& \qthree          (Q^{22}_\ell(z) Q_{bc} Q'_{b'c'} + R^{22}_\ell(z) Q_{bc} U'_{b'c'})                  \nn
\eea
To evaluate analagous sums containing $Y^C_{\ell m}$ or $Y^B_{\ell m}$, note that the
$90^\circ$ vector rotation 
$V_a \rightarrow -\epsilon_{ab}V^b$ sends
$Y^G_{\ell m} \rightarrow Y^C_{\ell m}$,
$X \rightarrow Y$,
$Y \rightarrow (-X)$; and the
$45^\circ$ tensor rotation 
$\Pi_{ab} \rightarrow (-\epsilon_{a}{}^{c}\Pi_{cb} - \epsilon_{b}{}^{c}\Pi_{ca})/2$ sends
$Y^E_{\ell m} \rightarrow Y^B_{\ell m}$,
$Q \rightarrow U$,
$U \rightarrow (-Q)$.
For example, the following sum will be used in Appendix \ref{atr}: 
\bea
&&  \sum_m Y^{G*}_{(\ell m)a}(\bx) Y^B_{(\ell m)bc}(\bx') = \frac{2\ell+1}{4\pi} \times  \label{esp8} \\
&&  \qthree (Q^{12}_\ell(z) X_a U'_{b'c'} - R^{12}_\ell(z) Y_a Q'_{b'c'})  \nn
\eea
It is obtained from the fifth equation in~(\ref{esp7}) by applying a $45^\circ$ rotation to the indices $bc$,
and no rotation to the index $a$.

As an application, using sums of the form (\ref{esp7}),
one can write down all correlation functions between Gaussian fields with spins $\le 2$ which
arise from parity-even power spectra:
{ \allowdisplaybreaks
\bea
\langle TT\rangle &=& \sum_\ell \frac{2\ell+1}{4\pi} C_\ell^{TT} P_\ell(z)                                      \label{esp9} \\
\langle XX\rangle &=& \sum_\ell \frac{2\ell+1}{4\pi} (C_\ell^{GG} Q^{11}_\ell(z) + C_\ell^{CC} R^{11}_\ell(z))  \nn   \\
\langle TX\rangle &=& -\sum_\ell \frac{2\ell+1}{4\pi} C_\ell^{TG} P_\ell^{01}(z)                                \nn   \\
\langle YY\rangle &=& \sum_\ell \frac{2\ell+1}{4\pi} (C_\ell^{GG} R^{11}_\ell(z) + C_\ell^{CC} Q^{11}_\ell(z))  \nn   \\
\langle TQ\rangle &=& -\sum_\ell \frac{2\ell+1}{4\pi} C_\ell^{TE} P_\ell^{02}(z)                                \nn   \\
\langle QQ\rangle &=& \sum_\ell \frac{2\ell+1}{4\pi} (C_\ell^{EE} Q^{22}_\ell(z) + C_\ell^{BB} R^{22}_\ell(z))  \nn   \\
\langle XQ\rangle &=& \sum_\ell \frac{2\ell+1}{4\pi} (C_\ell^{GE} Q^{12}_\ell(z) + C_\ell^{CB} R^{12}_\ell(z))  \nn   \\
\langle UU\rangle &=& \sum_\ell \frac{2\ell+1}{4\pi} (C_\ell^{EE} R^{22}_\ell(z) + C_\ell^{BB} Q^{22}_\ell(z))  \nn   \\
\langle YU\rangle &=& \sum_\ell \frac{2\ell+1}{4\pi} (C_\ell^{GE} R^{12}_\ell(z) + C_\ell^{CB} Q^{12}_\ell(z))  \nn
\eea
}
Here, correlation functions on the left-hand side are defined in the ``two-point'' basis from \S\ref{sn}; e.g.
$\langle XQ \rangle$ denotes the correlation between the $X_a$ component of a spin-1 field and the $Q'_{ab}$ component
of a spin-2 field, at points $\bx,\bx'$ with separation $z=(\bx\cdot \bx')$.

Correlation functions for parity-odd power spectra are given by:
\bea
\langle TY\rangle &=& -\sum_\ell \frac{2\ell+1}{4\pi} C_\ell^{TC} P_\ell^{01}(z)                                    \\
\langle TU\rangle &=& -\sum_\ell \frac{2\ell+1}{4\pi}  C_\ell^{TB} P_\ell^{02}(z)                              \nn  \\
\langle XY\rangle &=& \sum_\ell \frac{2\ell+1}{4\pi}  C_\ell^{GC} (Q_\ell^{11}(z) - R_\ell^{11}(z))            \nn  \\
\langle XU\rangle &=& \sum_\ell \frac{2\ell+1}{4\pi} (C_\ell^{GB} Q_\ell^{12}(z) - C_\ell^{CE} R_\ell^{12}(z)) \nn  \\
\langle YQ\rangle &=& \sum_\ell \frac{2\ell+1}{4\pi} (C_\ell^{CE} Q_\ell^{12}(z) - C_\ell^{GB} R_\ell^{12}(z)) \nn  \\
\langle QU\rangle &=& \sum_\ell \frac{2\ell+1}{4\pi} C_\ell^{EB} (Q_\ell^{22}(z) - R_\ell^{22}(z))             \nn
\eea
For spins 0 and 2, these results have already appeared in the literature \citep{Zpolexp,TOpolcf}, 
with the following explicit formulas for $P^{02}_\ell$, $Q^{22}_\ell$, $R^{22}_\ell$
used in place of the recursion relation (\ref{esp1}):
\begin{eqnarray}
P^{02}_\ell(z) &=& 2 \frac{        \frac{\ell z}{1-z^2} P_{\ell-1}(z)
                          - \left( \frac{\ell}{1-z^2} + \frac{\ell(\ell-1)}{2} \right) P_\ell(z)}
                           { [ (\ell-1)\ell(\ell+1)(\ell+2)]^{1/2}  }     \nn   \\
Q^{22}_\ell(z) &=& 2 \frac{   \frac{(\ell+2)z}{(1-z^2)} P^2_{\ell-1}(z)
                          - \left( \frac{\ell-4}{1-z^2} + \frac{\ell(\ell-1)}{2} \right) P_\ell^2(z)}
                             { (\ell-1)\ell(\ell+1)(\ell+2)  }            \nn   \\
R^{22}_\ell(z) &=& -4 \frac{  (\ell+2)P_{\ell-1}^2(z) - (\ell-1)zP_\ell^2(z)}
                             { (\ell-1)\ell(\ell+1)(\ell+2)(1-z^2)  }     \label{esp11}
\end{eqnarray}
(See also \citep{CLcf} for some spin-1 equations.)
The advantage of the present treatment is that it generalizes straightforwardly to all spins,
including spins higher than 2, and allows use of a single uniform recursion relation (\ref{esp1}) rather than
many explicit formulas of type (\ref{esp11}).
Our approach is similar to \citet{NgLiu}, who compute correlation functions covariantly for spins 0 and 2.

\section{Computing transfer matrices}
\label{atr}
There is still one ingredient missing from our formalism: an algorithm
for efficient calculation of the transfer matrices $K^{\pm pure}_{\ell\ell'}$,
which are needed to debias the power spectrum estimators (Eq.\ (\ref{ee1})).
We have found it more convenient to compute transfer matrices using
correlation functions than in harmonic space.
Since this approach has not been used previously in the literature, we
first illustrate the method for the simpler case of temperature
pseudo-$C_\ell$ estimators (\S\ref{str1}),
before treating the case of pure polarization estimators (\S\ref{str2}).
In \S\ref{str3}, we address implementational issues and show that the computational
cost of computing the transfer matrices is $\bigoh(\ell_{max}^3)$.

\subsection{Transfer matrices for temperature pseudo-$C_\ell$'s}
\label{str1}

In the temperature-only case, pseudo-$C_\ell$ estimators are constructed in a way which is completely
analagous to the polarization case (Eqs.\ (\ref{ee5})-(\ref{ee2})).  We briefly summarize the necessary definitions; 
details can be found in \citet{HGHCl}.

Pseudo multipoles and power spectra are defined by
\begin{eqnarray}
\widetilde T_{\ell m} &=& \sum_{\bx} T(\bx) W(\bx) Y_{\ell m}(\bx)^*    \\
\widetilde C_\ell^{TT}     &=& \frac{1}{2\ell+1} \sum_{m=-\ell}^\ell \widetilde T_{\ell m}^* \widetilde T_{\ell m}  \label{etr6}
\end{eqnarray}
Assuming no noise bias, the $(\ell_{max})$-by-$(\ell_{max})$ transfer matrix $K_{\ell\ell'}$ is defined by
\be
\langle \widetilde C_\ell^{TT} \rangle = \sum_{\ell'} K_{\ell\ell'} C^{TT}_{\ell'}
\ee
The purpose of this subsection is to compute $K_{\ell\ell'}$.  We first establish a key result: evaluation of a sum of the form
\be
\label{etr4}
S = \sum_{\bx \bx'} W(\bx) F(\bx\cdot \bx') W(\bx'),
\ee
where $F(z)$ is a polynomial in $z$ of degree $\le\ell_{max}$.
Expanding $F$ in Legendre polynomials, $F(z)=\sum_{\ell\le\ell_{max}} F_\ell P_\ell(z)$,
we evaluate the sum as follows:
\begin{eqnarray}
         S             &=& \sum_{\substack{\ell\le\ell_{max} \\ \bx \bx'}} W(\bx) W(\bx') F_\ell P_\ell(\bx\cdot \bx')  \label{etr5} \\
                       &=& \sum_{\substack{\ell\le\ell_{max} \\ \bx \bx' m}} \frac{4\pi}{2\ell+1} W(\bx) W(\bx') F_\ell Y_{\ell m}(\bx)^* Y_{\ell m}(\bx') \nn \\
                       &=& \sum_{\substack{\ell\le\ell_{max} \\ m}} W_{\ell m} W_{\ell m}^*  \frac{4\pi}{2\ell+1} F_\ell              \nn
\end{eqnarray}
It will be useful to rewrite this as an integral involving $F(z)$ instead of a sum involving $F_\ell$:
\be
S = 8\pi^2 \int_{-1}^1 dz\, \zeta^{WW}(z) F(z),   \label{etr3}
\ee
where
\be
\label{etr2}
\zeta^{WW}(z) \eqdef \frac{1}{4\pi} \sum_{\substack{\ell\le\ell_{max} \\ m}} |W_{\ell m}|^2 P_\ell(z)
\ee
can be thought of as an estimator of the correlation function of $W(x)$, pretending that $W(x)$ is a Gaussian
field on the full sky which is zero (by coincidence!) outside the survey region.

Armed with the key result (\ref{etr3}), the transfer matrix is calculated as follows.
By definition, the matrix element $K_{\ell \ell'}$ is the expectation value of $\widetilde C_\ell^{TT}$ (Eq.\ (\ref{etr6})), given the signal covariance
$\langle T(\bx) T(\bx') \rangle = (2\ell'+1)/(4\pi) P_{\ell'}(\bx\cdot \bx')$.  Therefore,
\begin{eqnarray}
K_{\ell\ell'}  &=& \frac{1}{2\ell+1} \sum_{\bx \bx' m} W(\bx) W(\bx') Y_{\ell m}(\bx) Y_{\ell m}(\bx')^*     \nn \\
                && \qtwo                        \left(\frac{2\ell'+1}{4\pi}\right) P_{\ell'}(\bx\cdot\bx')   \label{etr1}  \\
               &=& 2\pi \int_{-1}^1 dz\,\, \zeta^{WW}(z) P_\ell(z) \left(\frac{2\ell'+1}{4\pi}\right) P_{\ell'}(z).   \nn
\end{eqnarray}
We have now arrived at our desired expression for $K_{\ell\ell'}$.  We defer the issue of efficient evaluation of Eq.\ (\ref{etr1}) to
\S\ref{str3}.
For now, we note that Eq.\ (\ref{etr1}) is equivalent to the form commonly seen in the literature \citep[e.g.][Eq.\ A31]{MASTER},
\be
K_{\ell\ell'} = \frac{2\ell'+1}{4\pi} \sum_{\ell''m''} |W_{\ell''m''}|^2 \threej{\ell}{\ell'}{\ell''}{0}{0}{0}^2  \label{etr13}
\ee
by virtue of the definition (\ref{etr2}) of $\zeta^{WW}(z)$ and the identity
\be
\int_{-1}^1 dz\, P_\ell(z) P_{\ell'}(z) P_{\ell''}(z) = 2 \threej{\ell}{\ell'}{\ell''}{0}{0}{0}^2
\ee

\onecolumngrid

\subsection{The transfer matrix $K^{\pm pure}_{\ell\ell'}$}
\label{str2}

Computing the transfer matrix $K^{\pm pure}_{\ell\ell'}$ defined in Eq.\ (\ref{ec7}) is completely analagous to
the temperature-only case, but the bookkeeping is considerably more complicated owing to the presence
of spin-0, spin-1, and spin-2 weights $W(x)$, $W_a(x)$, $W_{bc}(x)$.
We will need to use the results in Appendix \ref{asp}, in which it is shown how to compute correlation functions
between fields of these spins from all parity-even power spectra.

Throughout this subsection, we will use the abbreviated notations
\begin{displaymath}
Y_{\ell m} \eqdef Y_{\ell m}(\bx) \qquad Y'_{\ell m} \eqdef Y_{\ell m}(\bx') \qquad z \eqdef (\bx\cdot\bx')
\end{displaymath}
in equations which contain a pair of points $\bx$, $\bx'$.

We first consider an analog of the ``key result'' Eq.\ (\ref{etr4}), in which $F$ is replaced by a 3-by-3 matrix which
contracts all combinations of spin-0, spin-1, and spin-2 weights:
\be
\label{etr7}
{\bf S} = \sum_{\bx \bx'} 
\left( \begin{array}{ccc} W(\bx) & W^a(\bx) & W^{bc}(\bx) \end{array} \right)
{\bf F}(\bx,\bx')
\left( \begin{array}{c} W(\bx) \\ W^{a'}(\bx') \\ W^{b'c'}(\bx') \end{array} \right).
\ee
where
\begin{displaymath}
{\bf F}(\bx,\bx') =
\left( \begin{array}{ccc}
      f^{WW}(z)     & f^{WX}(z)X'_{a'} & 2f^{WQ}(z)Q'_{b'c'}                           \\
      f^{WX}(z)X_a  &  f^{XX}(z)X_a X'_{a'} +  f^{YY}(z)Y_a Y'_{a'}
                            & 2f^{XQ}(z)X_a Q'_{b'c'} + 2f^{YU}(z)Y_a U'_{b'c'}        \\
  2f^{WQ}(z)Q_{bc}  & 2f^{XQ}(z)Q_{bc} X'_{a'}   + 2f^{YU}(z)U_{bc} Y'_{a'}
                            & 4f^{QQ}(z)Q_{bc} Q'_{b'c'} + 4f^{UU}(z)U_{bc} U'_{b'c'}
\end{array} \right)
\end{displaymath}
Note the bi-tensor structure of {\bf F}: each matrix entry carries a set of tensor indices which contract
with the weight functions in (\ref{etr7}) so that {\bf S} is a scalar.

We make the assumption that ${\bf F}$ is built from parity-even 
power spectra which are zero for multipoles larger than $\ell_{max}$.
(This is the analog of the assumption, from the preceding subsection, 
that $F(z)$ is a polynomial of degree $\le \ell_{max}$.)
To be precise, let $F_\ell^{WW}$, $F_\ell^{WG}$, $F_\ell^{WE}$, $F_\ell^{GE}$, 
$F_\ell^{EE}$, $F_\ell^{CC}$, $F_\ell^{CB}$, $F_\ell^{BB}$ be arbitrary power spectra.
Note that we denote the spin-0 ``field'' by $W$, and consider only parity-even combinations of fields.
Then ${\bf F}$ is assumed to be of the form,
\begin{eqnarray}
{\bf F}({\bf x},{\bf x}') &=&
\sum_{\substack{\ell \le \ell_{max} \\ m}}
\left( \begin{array}{ccc}
F_\ell^{WW} Y_{\ell m}^* Y_{\ell m}'  &  
        F_\ell^{WG} Y_{\ell m}^* Y_{(\ell m)a'}'^{G}  &  
        F_\ell^{WW} Y_{\ell m}^* Y_{(\ell m)b'c'}'^{E}  \\
F_\ell^{WG} Y_{(\ell m)a}^{G*} Y_{\ell m}'  &  
        F_\ell^{GG} Y_{(\ell m)a}^{G*} Y_{(\ell m)a'}'^{G}    & 
        F_\ell^{GE} Y_{(\ell m)a}^{G*} Y_{(\ell m)b'c'}'^{E}  \\
F_\ell^{WE} Y_{(\ell m)bc}^{E*} Y_{\ell m}'  &  
        F_\ell^{GE} Y_{(\ell m)bc}^{E*} Y_{(\ell m)a'}'^{G}    & 
        F_\ell^{EE} Y_{(\ell m)bc}^{E*} Y_{(\ell m)b'c'}'^{E}
\end{array} \right)  \nonumber  \\
&& \qquad\qquad +
\left( \begin{array}{ccc}
0  &  0  &  0  \\
0  &
        F_\ell^{CC} Y_{(\ell m)a}^{C*} Y_{(\ell m)a'}'^{C}    & 
        F_\ell^{CB} Y_{(\ell m)a}^{C*} Y_{(\ell m)b'c'}'^{B}  \\
0  &
        F_\ell^{CB} Y_{(\ell m)bc}^{B*} Y_{(\ell m)a'}'^{C}    & 
        F_\ell^{BB} Y_{(\ell m)bc}^{B*} Y_{(\ell m)b'c'}'^{B}
\end{array} \right).  \label{etr8}
\end{eqnarray}

The significance of the sum (\ref{etr7}), with {\bf F}-matrix of the form (\ref{etr8}), will appear shortly.
For now, we forge ahead and evaluate {\bf S}, using the method of the preceding subsection.
First, plugging the form (\ref{etr8}) for {\bf F} into the definition (\ref{etr7}) of {\bf S}, we get:
\begin{eqnarray}
{\bf S} &=& \sum_{\ell m} 
\left( \begin{array}{ccc}
W_{\ell m}  &  W_{\ell m}^{G}  &  W_{\ell m}^{E}
\end{array} \right)
\left( \begin{array}{ccc}
F_\ell^{WW} & F_\ell^{WG} & F_\ell^{WE}  \\
F_\ell^{WG} & F_\ell^{GG} & F_\ell^{GE}  \\
F_\ell^{WE} & F_\ell^{GE} & F_\ell^{EE}
\end{array} \right)
\left( \begin{array}{c}
W_{\ell m}^*  \\  W_{\ell m}^{G*}  \\ W_{\ell m}^{E*}
\end{array} \right)  \nonumber \\
&& \qquad\qquad +
\left( \begin{array}{cc}
W_{\ell m}^{C}  &  W_{\ell m}^{B}
\end{array} \right)
\left( \begin{array}{cc}
F_\ell^{CC} & F_\ell^{CB} \\
F_\ell^{BB} & F_\ell^{BB}
\end{array} \right)
\left( \begin{array}{c}
W_{\ell m}^{C*} \\ W_{\ell m}^{B*}
\end{array} \right)  \label{etr9}
\end{eqnarray}
This expression for ${\bf S}$ is the analog of Eq.\ (\ref{etr5}) from the preceding subsection.
Our goal is to get an expression analagous to Eq.\ (\ref{etr3}), in which the functions \{$f^{WW}(z)$, $f^{WX}(z)$, \ldots\}.
appear directly instead of the power spectra \{$F_\ell^{WW}$, $F_\ell^{WG}$, \ldots\}.
We first define correlation functions which are constructed from the weight functions
in the same way that one constructs correlation functions from parity-even power spectra (Eq.\ (\ref{esp9})):
\begin{eqnarray}
\zeta^{WW} &=& \frac{1}{4\pi} \sum_{\ell m} (W_{\ell m}^* W_{\ell m} P_\ell)   \label{etr10} \\
\zeta^{XX} = \frac{1}{4\pi} \sum_{\ell m} (W_{\ell m}^{G*} W_{\ell m}^G Q^{11}_\ell + W_{\ell m}^{C*} W_{\ell m}^C R^{11}_\ell)  &&
\zeta^{WX} = -\frac{1}{4\pi} \sum_{\ell m} (W_{\ell m}^* W_{\ell m}^G P_\ell^{01})   \nonumber \\
\zeta^{YY} = \frac{1}{4\pi} \sum_{\ell m} (W_{\ell m}^{G*} W_{\ell m}^G R^{11}_\ell + W_{\ell m}^{C*} W_{\ell m}^C Q^{11}_\ell)  &&
\zeta^{WQ} = -\frac{1}{4\pi} \sum_{\ell m} (W_{\ell m}^* W_{\ell m}^E  P_\ell^{02})  \nonumber \\
\zeta^{QQ} = \frac{1}{4\pi} \sum_{\ell m} (W_{\ell m}^{E*} W_{\ell m}^E Q^{22}_\ell + W_{\ell m}^{B*} W_{\ell m}^B R^{22}_\ell)  &&
\zeta^{XQ} = \frac{1}{4\pi} \sum_{\ell m} (W_{\ell m}^{G*} W_{\ell m}^E Q^{12}_\ell + W_{\ell m}^{C*} W_{\ell m}^B R^{12}_\ell)  \nonumber  \\
\zeta^{UU} = \frac{1}{4\pi} \sum_{\ell m} (W_{\ell m}^{E*} W_{\ell m}^E R^{22}_\ell + W_{\ell m}^{B*} W_{\ell m}^B Q^{22}_\ell)  &&
\zeta^{YU} = \frac{1}{4\pi} \sum_{\ell m} (W_{\ell m}^{G*} W_{\ell m}^E R^{12}_\ell + W_{\ell m}^{C*} W_{\ell m}^B Q^{12}_\ell)  \nonumber
\end{eqnarray}
Eq.\ (\ref{etr9}) can then be rewritten as an integral containing the functions \{$f^{WW}(z)$, $f^{WX}(z)$, \ldots\}:
\begin{eqnarray}
{\bf S} &=& 8\pi^2 \int_{-1}^1 dz\, \bigg[
                             \zeta^{WW}(z) f^{WW}(z)
                         + 2 \zeta^{WX}(z) f^{WX}(z)
                         + 2 \zeta^{WQ}(z) f^{WQ}(z)
                         + \zeta^{XX}(z) f^{XX}(z)    \nonumber \\
&& \qquad\qquad\qquad\qquad\qquad\qquad
                         + \,\zeta^{YY}(z) f^{YY}(z)
                         + \zeta^{YY}(z) f^{YY}(z)
                         + 2 \zeta^{XQ}(z) f^{XQ}(z)  \nonumber \\
&& \qquad\qquad\qquad\qquad\qquad\qquad
                         + \,2 \zeta^{YU}(z) f^{YU}(z)
                         + \zeta^{QQ}(z) f^{QQ}(z)
                         + \zeta^{UU}(z) f^{UU}(z) \bigg]  \label{etr11}
\end{eqnarray}
To show this, it is easiest to work backwards, substituting Eqs.\ (\ref{etr10}) into Eq.\ (\ref{etr11}), and ending up with Eq.\ (\ref{etr9}).
One writes $Q_\ell^{ss'}$, $R_\ell^{ss'}$ in terms of $P_\ell^{ss'}$ (Eq.\ (\ref{esp15})),
and uses the orthogonality relation for $P_\ell^{ss'}$ (Eq.\ (\ref{esp14})).

We have now arrived at our desired ``key result'': an expression for ${\bf S}$ in terms of the functions \{$f^{WW}(z)$, $f^{WX}(z)$, \ldots\}.
We now proceed to calculate the transfer matrix element $K^{+pure}_{\ell\ell'}$.
(The case of $K^{-pure}_{\ell\ell'}$ will be treated shortly.)
This matrix element is the expectation value $\langle \widetilde C^{BB,pure}_\ell \rangle$ given the signal covariance
\be
\langle \Pi^{de}({\bf x}) \Pi^{d'e'}({\bf x}') \rangle 
     = \left( \frac{2\ell'+1}{4\pi} \right) (R^{22}_{\ell'}(z) Q^{de} Q'^{d'e'} + Q^{22}_{\ell'}(z) U^{de} U'^{d'e'}).  \label{etr15}
\ee
A short calculation, using only the definition of $\widetilde C^{BB,pure}_\ell$ (Eqs.\ (\ref{ec4}), (\ref{ec9})), shows that
\be
\label{etr12}
K^{+pure}_{\ell\ell'} = \sum_{\bx \bx'} 
\left( \begin{array}{ccc} W(\bx) & W^a(\bx) & W^{bc}(\bx) \end{array} \right)
{\bf F}_{\ell\ell'}(\bx,\bx')
\left( \begin{array}{c} W(\bx) \\ W^{a'}(\bx') \\ W^{b'c'}(\bx') \end{array} \right)
\ee
where the {\bf F}-matrix is given by
\begin{eqnarray*}
{\bf F}_{\ell\ell'}(\bx,\bx') &=& 
\frac{2\ell'+1}{\pi(2\ell+1)}
\big[ R^{22}_{\ell'}(z) Q^{de} Q'^{d'e'} + Q^{22}_{\ell'}(z) U^{de} U'^{d'e'} \big] \quad \times \\
&& \qquad
  \sum_{m=-\ell}^\ell
  \left( \begin{array}{c}
     Y^{B*}_{(\ell m)de}  \\
       N'_\ell T_{dea}{}^{f} Y^{G*}_{(\ell m)f}   \\
       N_\ell T_{debc} Y^*_{\ell m}
  \end{array} \right)
  \left( \begin{array}{ccc}
     Y'^B_{(\ell m)d'e'}  &
       N'_\ell T_{d'e'a'}{}^{f'} Y'^{G}_{(\ell m)f'}  &
       N_\ell T_{d'e'b'c'} Y'_{\ell m}
  \end{array} \right)
\end{eqnarray*}
We will evaluate the right-hand side of Eq.\ (\ref{etr12}) using the key result (\ref{etr11}),
but first we simplify the {\bf F}-matrix using identities from Appendix \ref{asp}.
We show the details for the (1,2) matrix entry,
\be
{\bf F}^{(1,2)}_{\ell\ell'} = N'_\ell \frac{2\ell'+1}{\pi(2\ell+1)}
\big[ R^{22}_{\ell'}(z) Q^{de} Q'^{d'e'} + Q^{22}_{\ell'}(z) U^{de} U'^{d'e'} \big]
\sum_{m=-\ell}^\ell 
     Y^{B*}_{(\ell m)de}
     T_{d'e'a'}{}^{f'} 
     Y'^{G}_{(\ell m)f'}
\ee
To do the sum over $m$, we use identity (\ref{esp8}), obtaining
\be
{\bf F}^{(1,2)}_{\ell\ell'} =  N'_\ell
\frac{2\ell+1}{4\pi^2}
\big[ R^{22}_{\ell'}(z) Q^{de} Q'^{d'e'} + Q^{22}_{\ell'}(z) U^{de} U'^{d'e'} \big]
     T_{d'e'a'}{}^{f'} 
    ( Q_\ell^{12}(z) U_{de} X'_{f'} - R_\ell^{12}(z) Q_{de} Y'_{f'} )
\ee
It is then straightforward to do the index contractions and obtain:
\be
{\bf F}^{(1,2)}_{\ell\ell'} = -N'_\ell  \frac{2\ell'+1}{16 \pi^2}   (Q_\ell^{12} Q_{\ell'}^{22} + R_\ell^{12} R_{\ell'}^{22}) X'_{a'}
\ee
This procedure can be used to simplify the remaining entries of ${\bf F}_{\ell\ell'}$; when the dust has settled, one finds
\begin{eqnarray}
{\bf F}^{(1,1)}_{\ell\ell'} &=&  \frac{2\ell'+1}{16\pi^2} (Q_\ell^{22} Q_{\ell'}^{22} + R_\ell^{22} R_{\ell'}^{22})   \label{etr14}  \\
{\bf F}^{(1,3)}_{\ell\ell'} &=&  N_\ell \frac{2\ell'+1}{8\pi^2} P_\ell^{02} Q_{\ell'}^{22} Q'_{b'c'}                \nonumber \\
{\bf F}^{(2,2)}_{\ell\ell'} &=&  N^{'2}_\ell \frac{2\ell'+1}{16\pi^2} \bigg[
                                     (Q_\ell^{11} Q_{\ell'}^{22} + R_\ell^{11} R_{\ell'}^{22}) X_a X'_{a'}
                                   + (R_\ell^{11} Q_{\ell'}^{22} + Q_\ell^{11} R_{\ell'}^{22}) Y_a Y'_{a'} \bigg]      \nonumber \\
{\bf F}^{(2,3)}_{\ell\ell'} &=& -N_\ell N'_\ell \frac{2\ell'+1}{8\pi^2} P_\ell^{01} \big(
                                     Q_{\ell'}^{22} X_a Q'_{b'c'} + R_{\ell'}^{22} Y_a U'_{b'c'} \big)           \nonumber \\
{\bf F}^{(3,3)}_{\ell\ell'} &=&  N_\ell^2 \frac{2\ell'+1}{4\pi^2} P_\ell^{00} \big(
                                     Q_{\ell'}^{22} Q_{bc} Q'_{b'c'} + R_{\ell'}^{22} U_{bc} U'_{b'c'} \big)   \nonumber
\end{eqnarray}
We would like to apply the key result (\ref{etr11}) with {\bf F}-matrix given by this form, but first
there is an annoying technicality: our derivation of Eq.\ (\ref{etr11}) assumed that the {\bf F}-matrix was built from parity-even
power spectra, in the sense that Eq.\ (\ref{etr8}) is satisfied.
We claim that this is so for the {\bf F}-matrix in (\ref{etr14}), with $\ell_{max} = \ell+\ell'$.
To make this statement intuitively plausible, note that the {\bf F}-matrix in (\ref{etr14}) is a rotationally invariant, parity-even object
which is constructed by multiplying objects of spins $\ell$ and $\ell'$.
A formal proof can be given by writing $Q_\ell^{ss'}$, $R_\ell^{ss'}$ in terms of $P_\ell^{ss'}$ (Eq.\ (\ref{esp15})),
and using the product rule for $P_\ell^{ss'}$ (Eq.\ (\ref{esp13})).
This can be done one matrix entry at a time; again we supply the details only for the (1,2) matrix entry:
\begin{eqnarray}
{\bf F}^{(1,2)}_{\ell\ell'}(\bx,\bx')
&=& N'_\ell  \frac{2\ell'+1}{(4\pi)^2}   (Q_\ell^{12} Q_{\ell'}^{22} + R_\ell^{12} R_{\ell'}^{22}) X'_{a'}           \label{esp16} \\
&=& N'_\ell  \frac{2\ell'+1}{2(4\pi)^2} \big( P_\ell^{12}P_{\ell'}^{22} + P_\ell^{1,-2}P_{\ell'}^{2,-2} \big) X'_{a'}   \nonumber \\
&=& \sum_{|\ell-\ell'|\le\ell''\le\ell+\ell'} N'_\ell \frac{(2\ell'+1)(2\ell''+1)}{(4\pi)^2}
\threej{\ell}{\ell'}{\ell''}{1}{-2}{1} 
\threej{\ell}{\ell'}{\ell''}{2}{-2}{0} P^{10}_{\ell''} X'_{a'}   \nonumber  \\
&=& \sum_{\substack{|\ell-\ell'|\le\ell''\le\ell+\ell' \\ m''}} \bigg[ N'_\ell \frac{2\ell'+1}{4\pi} 
\threej{\ell}{\ell'}{\ell''}{1}{-2}{1} 
\threej{\ell}{\ell'}{\ell''}{2}{-2}{0}
\bigg] Y_{\ell'' m''}(\bx) Y^{G*}_{(\ell'' m'')a'}(\bx')   \nonumber
\end{eqnarray}
We have shown that ${\bf F}^{(1,2)}_{\ell\ell'}$ is of the form which appears on the RHS of Eq.\ (\ref{etr8}), with $\ell_{max} = \ell+\ell'$.

With this final technicality out of the way, we can use the key result (\ref{etr11}), with {\bf F}-matrix given by (\ref{etr14}), to evaluate
the right-hand side of Eq.\ (\ref{etr12}).
This gives the transfer matrix in the form:
\be
K^{+pure}_{\ell\ell'} = 2\pi \int_{-1}^1 dz\, \left( \frac{2\ell'+1}{4\pi} \right)
                                              \left( A_\ell(z) Q_{\ell'}^{22}(z) + B_\ell(z) R_{\ell'}^{22}(z) \right)   \label{esp18}
\ee
where the functions $A_\ell(z)$, $B_\ell(z)$ are defined by
\begin{displaymath}
A_\ell = \zeta^{WW} Q_\ell^{22} 
    - 2 N'_\ell \zeta^{WX} Q_\ell^{12}
    + 2  N_\ell \zeta^{WQ} P_\ell^{02}
  + N^{'2}_\ell (\zeta^{XX} Q_\ell^{11} + \zeta^{YY} R_\ell^{11})
- 2 N_\ell N'_\ell \zeta^{XQ} P_\ell^{01}
+   N_\ell^2 \zeta^{QQ} P_\ell^{00}
\end{displaymath}
\begin{displaymath}
B_\ell = \zeta^{WW} R_\ell^{22}
    - 2 N'_\ell \zeta^{WX} R_\ell^{12}
  + N^{'2}_\ell (\zeta^{XX} R_\ell^{11} + \zeta^{YY} Q_\ell^{11})
- 2 N_\ell N'_\ell \zeta^{YU} P_\ell^{01}
+   N_\ell^2 \zeta^{UU} P_\ell^{00}
\end{displaymath}
To compute $K^{-pure}_{\ell\ell'}$, one modifies the signal covariance (\ref{etr15}) by exchanging $Q_{\ell'}^{22} \leftrightarrow R_{\ell'}^{22}$;
this modification carries through to the end of the calculation and shows:
\be
K^{-pure}_{\ell\ell'} = 2\pi \int_{-1}^1 dz\, \left( \frac{2\ell'+1}{4\pi} \right)
                                              \left( A_\ell(z) R_{\ell'}^{22}(z) + B_\ell(z) Q_{\ell'}^{22}(z) \right)   \label{esp19}
\ee
Eqs.\ (\ref{esp18}) and (\ref{esp19}) are the main results of this appendix, and show how to compute the transfer matrices $K^{\pm pure}_{\ell\ell'}$
for any choice of weight functions.  

We emphasize that even though the derivation is lengthy, the final result is simple.
Each term in the integrals (\ref{esp18}), (\ref{esp19}) is a product of three functions: a generalized Legendre polynomial of degree $\ell'$,
a generalized Legendre polynomial of degree $\ell$, and a correlation function \{$\zeta^{WW}(z)$, $\zeta^{WX}(z)$, \ldots\} 
which depends only on the pixel weight functions (\ref{etr10}).
The final form of the transfer matrices is simpler in position space than in harmonic space, where the three-way multiplication
would be replaced by a sum involving 3j symbols, similar to Eq.\ (\ref{etr13}), but with many terms.

Finally, we discuss the case in which multipoles are binned into bandpowers.
In this case, the bandpower transfer matrix $K^{\pm}_{bb'}$ is related to $K^{\pm}_{\ell\ell'}$ by:
$K^{\pm}_{bb'} = P_{b\ell} K^{\pm}_{\ell\ell'} \bar P_{\ell'b'}$, where the matrices $P$ and $\bar P$ define the binning (\S\ref{se}).
For each bandpower, we define ``binned'' versions of the functions $A_\ell(z)$, $B_\ell(z)$, $Q^{22}_{\ell'}(z)$, $R^{22}_{\ell'}(z)$:
\begin{eqnarray}
A_b(z)    = \sum_\ell P_{b\ell} A_\ell(z)  & \qquad &   
B_b(z)    = \sum_\ell P_{b\ell} B_\ell(z)  \label{etr16} \\
Q_{b'}(z) = \sum_{\ell'} \left( \frac{2\ell'+1}{4\pi} \right) \bar P_{\ell'b'} Q^{22}_{\ell'}(z)  & \qquad &   
R_{b'}(z) = \sum_{\ell'} \left( \frac{2\ell'+1}{4\pi} \right) \bar P_{\ell'b'} R^{22}_{\ell'}(z) \nonumber
\end{eqnarray}
In terms of these, the bandpower transfer matrices are given by:
\begin{eqnarray}
K^{+pure}_{bb'} &=& 2\pi \int_{-1}^1 dz\, \big( A_b(z) Q_{b'}(z) + B_b(z) R_{b'}(z) \big)   \label{esp17}   \\
K^{-pure}_{bb'} &=& 2\pi \int_{-1}^1 dz\, \big( A_b(z) R_{b'}(z) + B_b(z) Q_{b'}(z) \big)   \nonumber
\end{eqnarray}

\twocolumngrid

\subsection{Efficient calculation of the transfer matrix integrals}
\label{str3}
We have now shown that the transfer matrices $K^{\pm}_{\ell\ell'}$ can be represented in integral form (\ref{esp17}).
In this subsection, we give an algorithm for evaluating the integrals to machine precision, whose running time is $\bigoh(\ell_{max}^3)$.
Here, $\ell_{max}$ is the largest value of $(\ell+\ell')$ for which a transfer matrix $K^{\pm}_{\ell\ell'}$ must be computed.
This is the same cost, within a constant factor, of evaluating the estimators once (\S \ref{sc2}).

The first observation is that, when computing correlation functions \{$\zeta^{WW}(z)$, $\zeta^{WX}(z)$, \ldots\} using Eq.\ (\ref{etr10}),
it is only necessary to sum over multipoles $\ell \le \ell_{max}$.
This is because, as argued at the end of the preceding section, the {\bf F}-matrix can be written in the form (\ref{etr8}), 
and by Eq.\ (\ref{etr9}), only multipoles of the weight functions with $\ell \le \ell_{max}$ contribute to {\bf S}.

The second observation is that the integrands in (\ref{esp17}) are polynomials of degree $\le 2\ell_{max}$.
This can be seen term-by-term after plugging in the definitions of $A_\ell(z)$, $B_\ell(z)$.
For some terms, such as $\zeta^{WW}(z) Q^{22}_{\ell}(z) Q^{22}_{\ell'}(z)$, all three factors are polynomials;
for others, such as $\zeta^{WX}(z) Q^{12}_{\ell}(z) Q^{22}_{\ell'}(z)$, the last is a polynomial and the first
two are polynomials times $\sqrt{1-z^2}$.  There are no terms with an odd number of $\sqrt{1-z^2}$ factors.

Because of this second observation, the integrals can be done exactly using Gauss-Legendre quadrature \citep[\S4.5]{NR}
with $(\ell_{max} + 1)$ points.  Our algorithm for evaluating the integrals is therefore given as follows.
First, we compute spherical harmonic transforms \{$W_{\ell m}$, $W_{\ell m}^G$, $W_{\ell m}^C$, $W_{\ell m}^E$, $W_{\ell m}^B$\} of the weight functions.
Second, we compute the correlation functions \{$\zeta^{WW}(z)$, $\zeta^{WX}(z)$, \ldots\} 
at each of the $(\ell_{max}+1)$ quadrature points, using Eq.\ (\ref{etr10}).
Third, we compute the $4N_{band}$ values \{$A_b(z)$, $B_b(z)$, $Q_b(z)$, $R_b(z)$\} at each quadrature point $z$, using (\ref{etr16}).
Fourth, we loop over bands $b$, $b'$, computing $K^{\pm pure}_{bb'}$ by doing the integrals in (\ref{esp17}) by Gauss-Legendre quadrature.

Let us consider the running time of each of these stages.
The first stage is $\bigoh(\ell_{max}^3)$, using fast spherical harmonic transforms.
The second stage can be done in time $\bigoh(\ell_{max}^2)$ by first summing over $m$ in Eq.\ (\ref{etr10})) at fixed $\ell$
and then summing over $\ell$ at fixed $z$.
The third stage is $\bigoh(N_{band} \ell_{max}^2)$, evaluating the generalized Legendre polynomials by recursion (\ref{esp1}).
The fourth stage is $\bigoh(N_{band}^2 \ell_{max})$ since Gauss-Legendre quadrature is $\bigoh(\ell_{max})$ once
everything has been precomputed at the quadrature points.
Putting this together, and noting that $N_{band} \le \ell_{max}$, the computational cost of computing the
transfer matrices $K^{\pm pure}_{\ell\ell'}$ is $\bigoh(\ell_{max}^3)$.

\section{Computing noise bias}
\label{anb}
In addition to the transfer matrix $K^{\pm pure}_{\ell\ell'}$, the estimators presented in this
paper also require computing noise bias terms $\widetilde N^{EE}_\ell$, $\widetilde N^{BB,pure}_\ell$
(Eq.\ (\ref{ec7})).  
For real experiments, which include such complications as $1/f$ noise, noise bias for all types
of pseudo-$C_\ell$ estimators must be computed by Monte Carlo.
Indeed, a practical advantage of the pseudo-$C_\ell$ framework is that unbiased estimators
can be constructed given only Monte Carlo simulations of the noise; no other representation
of the noise covariance is required.
However, for theoretical studies, it is convenient to have an exact formula for the noise bias
in simple cases.
In this appendix, we consider noise which is uncorrelated between pixels, and isotropic in each pixel,
but not necessarily homogeneous:
\be
\label{enb1}
\langle Q(\bx)Q(\bx) \rangle = \langle U(\bx)U(\bx) \rangle = \sigma(\bx)^2 \delta_{\bx\bx'}
\ee
Here, we represent the noise by its per-pixel RMS temperature $\sigma(\bx)$ (i.e., units $\mu$K
rather than $\mu$K-arcmin).

With noise covariance given by (\ref{enb1}), the noise bias defined in (\ref{ec7}) is given by
\bea
\widetilde N^{EE}_\ell &=& \sum_{\bx} \frac{\sigma^2(\bx)}{4\pi} W(\bx)^2            \\
\widetilde N^{BB,pure}_\ell &=& \sum_{\bx} \frac{\sigma^2(\bx)}{4\pi}
                                \bigg[ W(\bx)^2 + N^{'2}_\ell W_a(\bx) W^a(\bx)  \nn \\ 
             && \qtwo + 2 N^2_\ell W_{bc}(\bx) W^{bc}(\bx) \bigg]    \nn
\eea
This is derived starting from the definitions of $\widetilde E_{\ell m}$, $\widetilde B_{\ell m}$ 
(Eqs.\ (\ref{ee5}), (\ref{ec4}))
using identity (\ref{esp7}) from Appendix \ref{asp}.

\onecolumngrid

\section{Fisher matrix evaluation with azimuthal symmetry}
\label{afi}

In this appendix, we present the details of our method for fast exact evaluation of the Fisher matrix (\ref{ex11}),
in the case of inhomogeneous, but azimuthally symmetric, noise.
A similar method, in the context of CMB temperature, appeared in \citep{OSH}.
Since we only consider uncorrelated noise in this paper, the noise covariance can be written
\be
\label{efi1}
\big\langle Q({\bf x}), Q({\bf x'}) \big\rangle = \big\langle U({\bf x}), U({\bf x'}) \big\rangle = \eta(\theta)^2 \delta^{(2)}({\bf x}-{\bf x}'),
\ee
where $\eta(\theta)$ is arbitrary.
If we change variables from $\{Q,U\}$ to $\Pi^{\pm} = (Q\pm iU)$, and Fourier transform in the azimuthal coordinate $\varphi$, by defining
\be
\widetilde \Pi_m^\pm(\theta) = \int_0^{2\pi} d\varphi\, [Q\pm iU](\theta,\varphi) e^{im\varphi}
\ee
then the noise covariance (\ref{efi1}) is still diagonal (in both $m$ and $\theta$):
\be
\label{efi2}
\big\langle \Pi^{+*}_m(\theta) \Pi^+_{m'}(\theta') \big\rangle =
\big\langle \Pi^{-*}_m(\theta) \Pi^-_{m'}(\theta') \big\rangle =
4\pi \frac{\eta(\theta)^2}{\sin(\theta)} \delta(\theta-\theta') \delta_{mm'}
\ee
The point of this change of variables is that the signal covariance is also diagonal in $m$ (but still dense in $\theta$).
Using the results of \citep{ZSspins}, one can show that the signal covariance is given by:
\begin{eqnarray}
&& \left\langle
  \left( \begin{array}{c}  \widetilde \Pi_m^{+*}(\theta)  \\ \widetilde \Pi_m^{-*}(\theta)  \end{array} \right)
  \left( \begin{array}{cc} \widetilde \Pi_{m'}^{+}(\theta')   &  \widetilde \Pi_{m'}^{-}(\theta')   \end{array} \right)
\right\rangle = 4\pi^2 \delta_{mm'} \quad \times \label{efi3}   \\
&& \qquad\qquad \sum_\ell \left( \begin{array}{cc}
    (C_\ell^{EE} + C_\ell^{BB})    ({}_{-2}Y_{\ell m}(\theta,0))  ({}_{-2}Y_{\ell m}(\theta',0)))    &
       (C_\ell^{EE} - C_\ell^{BB}) ({}_{-2}Y_{\ell m}(\theta,0))  ({}_{2}Y_{\ell m}(\theta',0)) \\
    (C_\ell^{EE} - C_\ell^{BB})    ({}_{2}Y_{\ell m}(\theta,0))   ({}_{-2}Y_{\ell m}(\theta',0))     &
       (C_\ell^{EE} + C_\ell^{BB}) ({}_{2}Y_{\ell m}(\theta,0))   ({}_{2}Y_{\ell m}(\theta',0))
\end{array} \right)  \nonumber
\end{eqnarray}
Eqs.\ (\ref{efi2}) and (\ref{efi3}) can be used to efficiently calculate the Fisher matrix
\be
\label{efi4}
F_{bb'} = \frac{1}{2} \Tr( {\bf S}_b ({\bf S}_0 + {\bf N})^{-1} {\bf S}_{b'} ({\bf S}_0 + {\bf N})^{-1} )
\ee
For concrete calculation, the continuous coordinate $\theta$ is replaced with a set of $N$ 
equally spaced values $\theta_1$, $\theta_2$, $\cdots$, $\theta_N$, with spacing $\Delta\theta$.
The $(2N)$-by-$(2N)$ noise and signal covariances of the variables 
\{$\Pi^+_m(\theta_1)$, $\cdots$, $\Pi^+_m(\theta_N)$, $\Pi^-_m(\theta_1)$, $\cdots$, $\Pi^-_m(\theta_N)$\}
are then given by
\be
{\bf N}^{(m)} = \frac{4\pi}{\Delta\theta}
\left( \begin{array}{cc}
\frac{\eta(\theta_i)^2}{\sin(\theta_i)}\delta_{ij} & 0   \\
0  & \frac{\eta(\theta_i)^2}{\sin(\theta_i)}\delta_{ij}
\end{array} \right)
\ee
\begin{displaymath}
{\bf S}^{(m)} = 4\pi^2 \sum_\ell
\left( \begin{array}{cc}
    (C_\ell^{EE} + C_\ell^{BB})  \,    {}_{-2}Y_{\ell m}(\theta_i,0)  \,    {}_{-2}Y_{\ell m}(\theta_j,0)    &
    (C_\ell^{EE} - C_\ell^{BB})  \,    {}_{-2}Y_{\ell m}(\theta_i,0)  \,    {}_{2}Y_{\ell m}(\theta_j,0) \\
    (C_\ell^{EE} - C_\ell^{BB})  \,    {}_{2}Y_{\ell m}(\theta_i,0)   \,    {}_{-2}Y_{\ell m}(\theta_j,0)    &
    (C_\ell^{EE} + C_\ell^{BB})  \,    {}_{2}Y_{\ell m}(\theta_i,0)   \,    {}_{2}Y_{\ell m}(\theta_j,0)
\end{array} \right)
\end{displaymath}
The Fisher matrix (\ref{efi4}) is given by summing over $m$ and tracing over $\theta$:
\begin{eqnarray}
F_{bb'} &=& \frac{1}{2}\sum_{m=-m_{max}}^{m_{max}} \Tr
          ({\bf S}_b^{(m)} ({\bf S}_0^{(m)} + {\bf N})^{-1} {\bf S}_{b'}^{(m)} ({\bf S}_0^{(m)} + {\bf N})^{-1})  \\
        &=& \frac{1}{2} \Tr ({\bf S}_b^{(0)} ({\bf S}_0^{(0)} + {\bf N})^{-1} {\bf S}_{b'}^{(0)} ({\bf S}_0^{(0)} + {\bf N})^{-1}) +
          \sum_{m=1}^{m_{max}} \Tr
          ({\bf S}_b^{(m)} ({\bf S}_0^{(m)} + {\bf N})^{-1} {\bf S}_{b'}^{(m)} ({\bf S}_0^{(m)} + {\bf N})^{-1})  \nonumber
\end{eqnarray}
where ``Tr'' is a $(2N)$-by-$(2N)$ trace.
Evaluating the spin harmonics by recursion in $\ell$,
the computational cost of each Fisher matrix element is $\bigoh(N_\theta^3 m_{max})$, versus $\bigoh(N_{pix}^3)$ for brute
force calculation without exploiting azimuthal symmetry.
(For the mock surveys studied in this paper, with Healpix resolution $N_{side} = 256$, one has $N_\theta \sim 70$
and $N_{pix} \sim 10^4$.)
%
%
%
%
\twocolumngrid
\end{document}